\documentclass[aps,prd,reprint,superscriptaddress]{revtex4-2}
\usepackage{amsmath,amssymb,graphicx}
\usepackage{xcolor}
\usepackage{needspace}
\usepackage[normalem]{ulem}
\usepackage[hidelinks]{hyperref}
\hypersetup{pdftitle={Kerr QNM Convergence Region}}
\makeatletter
\renewcommand{\frontmatter@abstractwidth}{468pt}
\makeatother

\newcommand{\pr}{\mathrm{pr}}
\newcommand{\ob}{\mathrm{ob}}
\newcommand{\rbpr}{r_{\rm b}^{\pr}}
\newcommand{\rbob}{r_{\rm b}^{\ob}}
\newcommand{\IN}{\mathrm{IN}}
\newcommand{\UP}{\mathrm{UP}}
\newcommand{\dd}{\mathrm{d}}
\newcommand{\Li}{\operatorname{Li}}
\newcommand{\Imn}{\operatorname{Im}}
\newcommand{\Ren}{\operatorname{Re}}

\definecolor{DCDarkOrange}{HTML}{A34700}

\begin{document}
\title{Convergence of the quasinormal-mode expansion of the Kerr Green function}
\author{Davide Ciattaglia}
\affiliation{SISSA, Via Bonomea 265, 34136 Trieste, Italy}
\affiliation{INFN Sezione di Trieste, Via Valerio 2, 34127 Trieste, Italy}
\affiliation{IFPU -- Institute for Fundamental Physics of the Universe, Via Beirut 2, 34014 Trieste, Italy}
\author{Adrien Kuntz}
\affiliation{CENTRA, Departamento de F\'{\i}sica, Instituto Superior T\'ecnico -- IST, Universidade de Lisboa -- UL, Avenida Rovisco Pais 1, 1049-001 Lisboa, Portugal}
\author{Enrico Barausse}
\affiliation{SISSA, Via Bonomea 265, 34136 Trieste, Italy}
\affiliation{INFN Sezione di Trieste, Via Valerio 2, 34127 Trieste, Italy}
\affiliation{IFPU -- Institute for Fundamental Physics of the Universe, Via Beirut 2, 34014 Trieste, Italy}
\date{\today}
\begin{abstract}
Black-hole ringdown is usually modeled as a linear superposition of quasinormal modes, and fits to gravitational-wave observations increasingly include high overtones. However, the overtone sum does not necessarily converge: overtones are more strongly damped and thus decay faster in time, but their amplitudes in the Green function (the signal at an observer produced by an impulse at a source point) grow, so the mode description may break down before a finite time. We show that for a rotating (Kerr) black hole this time has a clear geometric meaning. For highly damped modes, the amplitudes grow exponentially with the optical distance from the source and  the observer to a characteristic ``bounce'' radius close to the black hole,
defined as the location where the mode amplitude is smallest. 
For one family of modes, the amplitudes  are also suppressed the farther the source and observer sit from the equatorial plane.  Rotation therefore makes the onset of convergence depend
on direction as well as distance, extending the geometric
picture from bounce radii to bounce surfaces. Combining these effects gives a sufficient condition for absolute convergence. For spin $a/M = 0.7$, we map how this time varies with source and observer position, and amplitudes computed numerically reproduce the predicted spatial dependence to better than one percent. These results set geometric limits on when
the quasinormal-mode sums converge.
\end{abstract}
\maketitle
\raggedbottom
\clubpenalty=10000
\widowpenalty=10000

\section{Introduction}\label{s:intro}
When two black holes merge, the remnant settles down to a stationary Kerr black hole by emitting gravitational waves. The final part of this signal, the ringdown, is well described by a superposition of quasinormal modes (QNMs): damped oscillations whose complex frequencies are, in general relativity, fixed entirely by the mass and spin of the remnant~\cite{Kokkotas:1999bd,Nollert:1999ji,Berti:2009kk,Berti:2025hly}. Measuring more than one such frequency therefore tests whether the remnant is a Kerr black hole, the idea behind black-hole spectroscopy~\cite{Dreyer:2003bv,Berti:2005ys,CalderonBustillo:2020rmh}. Such tests are expected to become routine with next-generation ground-based and space-based detectors~\cite{Berti:2016lat}, and precise measurements of several modes can constrain deviations from the Kerr multipole structure and reconstruct the perturbation potential near the light ring~\cite{Dey:2022pmv,Volkel:2022aca,Volkel:2022khh}. Ringdown tests have accompanied gravitational-wave astronomy since GW150914~\cite{LIGOScientific:2016lio,LIGOScientific:2021sio,LIGOScientific:2026wpt}, and the loud event GW250114 has now resolved the dominant quadrupolar mode together with its first overtone~\cite{LIGOScientific:2025rid}.

Overtones, i.e. the more strongly damped modes with the same angular indices, have become central to this program. Including several of them allows linear QNM models to reproduce numerical-relativity waveforms starting near the peak of the signal~\cite{Giesler:2019uxc}, and overtone fits have been applied to observational data. For GW150914, the first overtone was initially reported at $3.6\sigma$ in a fit starting at the amplitude peak~\cite{Isi:2019aib}. Subsequent analyses reached differing conclusions, with the evidence depending on the start time, the noise estimate, the statistical methods, and the treatment of sky-location uncertainties~\cite{Cotesta:2022pci,Isi:2022mhy,Finch:2022ynt,Ma:2023vvr,Isi:2023nif,Carullo:2023gtf,Wang:2023ljx,Correia:2023bfn,Crisostomi:2023tle,Dey:2026btx}. For GW190521, evidence for a subdominant $(\ell,m,n)=(3,3,0)$ mode has been reported~\cite{Capano:2021etf}, with weaker or differently interpreted support in other analyses~\cite{Siegel:2023lxl,Dey:2026btx}. For the much louder GW250114, the first overtone is resolved in post-peak data~\cite{LIGOScientific:2025rid,LIGOScientific:2025wao}; this detection is robust to the treatment of extrinsic parameters, but its evidence fades within about $10M$ after the peak~\cite{Dey:2026btx}. When many overtones are included, the preferred start time depends on their number, and the amplitudes of the highest overtones become strongly correlated~\cite{Coleman:2025ipc}. For future space-based observations, truncating the mode content biases the inferred remnant parameters, and the number of modes required depends on the source mass and redshift~\cite{Capuano:2025kkl}. Whether  early-time fits capture physical overtones, or instead absorb nonlinear effects, transients and the still-active source, remains debated~\cite{Baibhav:2023clw,Nee:2023osy,Mitman:2022qdl,Cheung:2022rbm,DeAmicis:2025xuh}. The overtone frequencies are also unstable under small-scale perturbations of the effective potential~\cite{Jaramillo:2020tuu}. Underlying these questions is a more basic one: where, and from what time onward, does a sum over overtones represent the linear response of a black hole at all?

In linear theory, the response to any source follows from the retarded Green function. Following Leaver~\cite{leaver1986}, its frequency-domain representation can be split into QNM poles, a branch cut responsible for late-time power-law tails, and high-frequency contributions that dominate the prompt response~\cite{leaver1986,Andersson:1996cm}. Unlike the normal modes of a closed system, QNMs form a complete set only for special classes of open systems~\cite{Ching:1998mxl}; for black holes the QNM sum must be supplemented by the other contributions, and it does not need to converge~\cite{Andersson:1996cm,Nollert:1999ji}. The non-QNM contributions can themselves acquire a mode-like description. In Schwarzschild, the branch cut can be written as a convergent sum of de~Sitter QNMs in the limit of vanishing cosmological constant~\cite{Arnaudo:2025uos}. Similarly, for a black hole whose mass changes at a constant rate, the perturbations can still be treated in the frequency domain~\cite{Capuano:2024qhv}, and the spectrum contains purely imaginary modes in addition to the light-ring QNMs, which reduce to the familiar power-law tail in the Schwarzschild limit~\cite{Capuano:2026tjy}. 

Each QNM term is a damped exponential multiplied by an amplitude that depends on the source and observer positions. 

In Schwarzschild, the excitation factors,
which depend only on the black hole,
vary as a power of the overtone number at large damping,
but the amplitudes at fixed source and observer grow
exponentially with damping through the radial wave
solutions~\cite{Andersson:1996cm,bertiExc2006}. For a source and an observer
far from the hole, the QNM sum therefore converges once the elapsed time exceeds $r_*+r_*'$, the sum of their tortoise coordinates~\cite{leaver1986,Andersson:1996cm,bertiExc2006}. Faster temporal decay of individual modes does not by itself ensure that the sum converges. Convergence instead results from a competition between temporal decay and amplitude growth, and holds only in part of spacetime.

For Schwarzschild black holes this competition has recently acquired a simple geometric picture. Arnaudo \emph{et al.} showed that the QNM sum of the retarded Green function converges at late times, a complementary Matsubara sum converges at early times, and the two regions are separated by a light cone that scatters from the black-hole potential~\cite{Arnaudo:2025uos}. The scattering point or bounce radius lies at $r_*=0$ in standard tortoise conventions, and its location was traced to a complex-time ``bouncing singularity'', associated with a null geodesic that bounces off the black-hole singularity~\cite{Arnaudo:2026tcy}. De Amicis \emph{et al.} confirmed the role of this bounce radius 
 (defined as the radius where the amplitude is smallest for highly damped modes) 
in the QNM excitation by a plunging particle: while the particle is outside it, successive overtones are driven in counter-phase and interfere destructively, whereas after it crosses, they oscillate in phase and the QNM sum converges rapidly~\cite{DeAmicis:2025xuh,DeAmicis:2026wqd}. For a particle plunging from the innermost stable circular orbit, a QNM sum with constant amplitudes converges only after a start time about $9.4M$ after the peak of the quadrupolar mode, set by the advanced time at which the particle reaches the singularity~\cite{Kuntz:2026crc}. For Kerr, Oshita identified the QNM excitation radius from the Stokes geometry of WKB solutions, recovering $r\simeq2.557M$ in the Schwarzschild limit~\cite{oshita2026}, and Su \emph{et al.} decomposed the time-domain Kerr Green function into QNM, direct and tail parts, with split times that depend on the spin and on the emitter and receiver positions~\cite{su2026}.

Rotation changes the problem qualitatively. In Kerr, the radial and angular dependence of a perturbation are coupled through the complex frequency: the angular functions are spin-weighted spheroidal harmonics whose shape depends on $a\omega$~\cite{teukolsky1973,bertiAngular2006}. This separation of variables is itself special to Kerr; while it has been very recently extended to several other spacetimes~\cite{Hintz:2026jiw,Santos:2026pbu,Dias:2026zri}, it is lost in generic stationary, axisymmetric geometries~\cite{Ghosh:2023etd}.  At large damping the angular eigenvalue follows one of two distinct asymptotic behaviors: the ``normal'' prolate behavior and an ``anomalous'' one that mirrors the oblate limit~\cite{vickersCook2022} (see also Refs.~\cite{bertiAngular2006,keshetBenMeir2014}). We refer to the corresponding modes as prolate and oblate. The two behaviors produce different radial propagation, and hence different bounce radii. They also produce very different angular functions: prolate modes become concentrated in an ever narrower belt around the equatorial plane, whereas oblate modes oscillate across the sphere. The amplitude of a Kerr QNM therefore depends on the polar angles of the source and observer as well as on their radii, and the convergence time inherits this direction dependence. Here we focus on the angular weight at the source and observer, on the competition between the two families, and on a direct test against numerically computed amplitudes.

Our main results are as follows. At large damping, the
amplitudes have an exponential growth rate whose radial
part is the sum
of the light-travel times from the corresponding bounce radius
to the source and observer radii. These times correspond
to special null rays: along the rotation axis for oblate
modes and in the equatorial plane, with zero axial angular
momentum, for prolate modes. Prolate angular localization
subtracts an explicit term that grows with the angular
distance of the source and observer from the equator,
whereas oblate modes carry no such penalty. Together,
the radial and angular contributions define
\textit{bounce surfaces}, whose radii depend on polar
angle in the prolate sector.
The prolate rate combines the passage times of null
fronts at the source and observer, describing an apparent
bounce of the whole wavefront when the two fronts approach
and leave the same surface (Sec.~\ref{s:geometry}).
The larger of the two rates gives a sufficient time for
absolute convergence of the QNM sums at fixed angular
indices $(\ell,m)$.
This threshold need not coincide with the first causal
arrival: the pole sums can converge before any signal
reaches the observer (Sec.~\ref{s:competition}).
Growth rates extracted from numerical amplitudes, which
include radial wave solutions, the angular projector and
the pole normalization, reproduce the predicted spatial
dependence to better than one percent, and complex ratios
of the numerical amplitudes reproduce the predicted phases
without fitting. In the zero-spin limit both families
recover the Schwarzschild bounce at $r_*=0$.

For a physical source, this time must be reached for every source element that contributes, so the overtone description is not guaranteed to hold while an extended source is still active near the black hole. Rotation makes this time depend on direction: prolate overtones are suppressed for sources and observers away from the equatorial plane, but oblate overtones are not. Our analysis concerns sources and observers at finite radii. Relating it to waveforms measured at large distances, and to the start time of ringdown fits, requires additional steps beyond this work.

The paper is organized as follows. Section~\ref{s:method} sets up the Teukolsky equation, the retarded Green function and its QNM amplitudes. Section~\ref{s:wkb} derives the radial and angular growth rates at large damping. Section~\ref{s:competition} compares the two families and states the resulting convergence condition. 
Section~\ref{s:geometry} interprets the growth rates
in terms of null rays, wavefronts and bounce surfaces.
Technical details are collected in the appendices. Throughout the article we use units $c=G=1$, and set the black-hole (BH) mass to $M=1$.

\section{Methodology}\label{s:method}

We construct the retarded Green function, which gives the response to an impulse at a source point. Its poles determine the QNM frequencies, and their residues determine the amplitudes entering the overtone sum. For gravitational perturbations of Kerr, we use the Teukolsky equation with spin weight $s=-2$~\cite{teukolsky1973}. For a black hole of mass $M$ and angular momentum $Ma$, its field variable is ${}_{-2}\Psi=(r-ia\cos\theta)^4\delta\psi_4$, where $\delta\psi_4$ is the perturbed Weyl-curvature component in the Kinnersley tetrad.

Separation lets us build the response one component at a time. In Boyer--Lindquist coordinates $x=(t,r,\theta,\phi)$, we expand in time and azimuth with complex frequency $\omega=\omega_{\rm R}+i\omega_{\rm I}$ and integer azimuthal number $m$. For each $\omega,m$ we expand the angular dependence in regular eigenfunctions labeled by $\ell$. One component has the form
\begin{equation}
 {}_s\Psi_{\ell m\omega}(x)
 =e^{-i\omega t+im\phi}\,
 {}_s R_{\ell m\omega}(r){}_s S_{\ell m\omega}(\theta)
 \label{e:separation}
\end{equation}
At this stage $\omega$ is a general frequency of the transform. We will find the QNM frequencies by locating the poles of the Green function. The angular equation determines the spheroidal harmonic ${}_s S_{\ell m\omega}$ and its eigenvalue ${}_s A_{\ell m}(a\omega)$. This eigenvalue enters the equation for ${}_s R_{\ell m\omega}$, so the angular pattern also affects radial propagation. Let $\Delta=r^2-2r+a^2$, whose zeros $r_\pm=1\pm\sqrt{1-a^2}$ are the horizon radii. The radial equation uses the combinations
\begin{align*}
 K_{m\omega}&=(r^2+a^2)\omega-am,\\
 {}_s\lambda_{\ell m\omega}&={}_s A_{\ell m}(a\omega)+a^2\omega^2-2am\omega
\end{align*}
The separated Teukolsky equations are
\begin{align}
 &\Delta^{-s}\partial_r
       \big(\Delta^{s+1}\partial_r {}_s R_{\ell m\omega}\big)
       \nonumber\\
 &+\bigg[\frac{K_{m\omega}^2-2is(r-1)K_{m\omega}}{\Delta}
       \nonumber\\
 &\qquad+4is\omega r-{}_s\lambda_{\ell m\omega}\bigg]
          {}_s R_{\ell m\omega}=0\label{e:TeukR}\\
 &\frac{1}{\sin\theta}\partial_\theta
      \big(\sin\theta\,\partial_\theta {}_s S_{\ell m\omega}\big)
       \nonumber\\
 &+\bigg[(a\omega)^2\cos^2\theta-2a\omega s\cos\theta
       +s+{}_s A_{\ell m}(a\omega)\nonumber\\
 &\qquad-\frac{(m+s\cos\theta)^2}{\sin^2\theta}
       \bigg]{}_s S_{\ell m\omega}=0
 \label{e:TeukS}
\end{align}
Once the angular eigenvalue is known, Eq.~\eqref{e:TeukR} determines how the disturbance travels from a source at radius $r'$ and time $t'$ to an observer at $r,t$. 
We define two homogeneous solutions of Eq.~\eqref{e:TeukR}, ${}_s R^{\IN}_{\ell m\omega}$ and ${}_s R^{\UP}_{\ell m\omega}$, respectively ingoing at the horizon and outgoing at infinity:
\begin{align}
\label{eq:RH_asymp}
    {}_s R^{\IN}_{\ell m\omega} &\to \Delta^{-s} e^{-i (\omega - m \Omega_\mathrm{H}) r_*}\ , & r \rightarrow r_+ \ ,\\
    \label{eq:Rinf_asymp}
   {}_s R^{\UP}_{\ell m\omega} &\to r^{-1-2s} e^{i \omega r_*} \ ,& r  \rightarrow \infty \ .
\end{align}
where $\Omega_{\rm H}=a(2 r_+)^{-1}$ is the horizon angular velocity and $r_*$ is the tortoise coordinate,
\begin{equation}\label{eq:tortoise}
    r_*=r + \frac{2r_+}{r_+-r_-} \log\left(\frac{r-r_+}{2}\right) -  \frac{2r_-}{r_+-r_-} \log\left(\frac{r-r_-}{2}\right).
\end{equation}
In frequency space the point source at $r'$ fixes a unit jump in $\Delta^{s+1}\partial_rG_{\rm rad}$ across $r'$. The radial Green function is
\begin{equation*}
 G_{\rm rad}(\omega;r,r')=
 \frac{{}_s R^{\IN}_{\ell m\omega}(r_<){}_s R^{\UP}_{\ell m\omega}(r_>)}
      {{}_s\widehat W_{\ell m}(\omega)}
\end{equation*}
where $r_<=\min(r,r')$ and $r_>=\max(r,r')$ and the Wronskian ${}_s\widehat W_{\ell m}(\omega)$ is defined by
\begin{align}
 {}_s\widehat W_{\ell m}(\omega)
 =\Delta^{s+1}\big[
 &{}_s R^{\IN}_{\ell m\omega}\,\partial_r{}_s R^{\UP}_{\ell m\omega}
 \nonumber\\[-2pt]
 &-(\partial_r{}_s R^{\IN}_{\ell m\omega}){}_s R^{\UP}_{\ell m\omega}
 \big]
 \label{e:wronsk}
\end{align}
The radial equation~\eqref{e:TeukR} makes ${}_s\widehat W_{\ell m}$ independent of $r$, although it varies with $\omega$. On the other hand, the full Green function also involves an angular projector $ {}_s\Pi_{\ell m\omega}(\theta,\theta')$~\cite{su2026}, where the source is located at $\theta'$ and the observer at $\theta$:

\Needspace{5\baselineskip}
\begin{align} \label{eq:angularProj}
 {}_s\Pi_{\ell m\omega}(\theta,\theta')
 &=\frac{{}_s S_{\ell m\omega}(\theta){}_s S_{\ell m\omega}(\theta')}
 {{}_s N_{\ell m\omega}}, \\
 {}_s N_{\ell m\omega}
 &=\int_0^\pi {}_s S_{\ell m\omega}(\vartheta)^2
                       \sin\vartheta\,\dd\vartheta
 \label{e:norm}
\end{align}
The angular projector ${}_s\Pi_{\ell m\omega}$ is independent of the normalization of ${}_s S_{\ell m\omega}$. 
For an impulse at $x'=(t',r',\theta',\phi')$, the product of this projector and the radial Green function gives the contribution at fixed $\ell,m,\omega$. Summing over angular indices and inverting the frequency transform gives the retarded response at $x$,
\begin{align}
 {}_s G^{\rm ret}(x,x')={}&\frac{1}{(2\pi)^2}
 \sum_{\ell,m}\int_{\mathcal C}\dd\omega\,
 e^{-i\omega(t-t')+im(\phi-\phi')}\nonumber\\
 &\times{}_s\Pi_{\ell m\omega}(\theta,\theta')
 \frac{{}_s R^{\IN}_{\ell m\omega}(r_<){}_s R^{\UP}_{\ell m\omega}(r_>)}
 {{}_s\widehat W_{\ell m}(\omega)}
 \label{e:retardedtransform}
\end{align}
With the convention $e^{-i\omega t}$, the inversion contour $\mathcal C$ runs from left to right along a horizontal line in an analytic region of the upper half-plane. To find poles below the real axis, we continue both radial solutions in $\omega$, following the angular eigenvalue ${}_s A_{\ell m}(a\omega)$ that enters their equation.

\subsection{Quasinormal modes and convergence}

At a QNM frequency ${}_s\omega_{\ell mn}$ the weighted Wronskian
${}_s\widehat W_{\ell m}$ vanishes. The horizon-ingoing solution is then
also outgoing at infinity. We consider damped modes with overtone index
$n$ and $\omega_{{\rm I},n}<0$. At a simple zero of $\widehat W$ with
a regular angular projector, the frequency-domain residue is
\begin{equation}
 {}_s\mathcal R_{\ell mn}
 =\left.
   \frac{{}_s\Pi_{\ell m\omega}(\theta,\theta')\,
         {}_s R^{\IN}_{\ell m\omega}(r_<){}_s R^{\UP}_{\ell m\omega}(r_>)}
        {\dd{}_s\widehat W_{\ell m}/\dd\omega}
  \right|_{{}_s\omega_{\ell mn}}
 \label{e:mainres}
\end{equation}
The denominator is the total frequency derivative along $\omega$, taking into account the dependence of the angular
eigenvalue ${}_s A_{\ell m}(a\omega)$ on the frequency. For $t>t'$ we deform the frequency
contour toward the lower half-plane to extract the QNM contributions~\cite{leaver1986}.
Each pole contributes a damped exponential. The clockwise orientation
gives $-2\pi i$ per residue. With the Fourier normalization $(2\pi)^{-2}$
in Eq.~\eqref{e:retardedtransform}, the poles at fixed $s,\ell,m$
contribute
\begin{equation}
 \begin{aligned}
 {}_s G^{\rm poles}_{\ell m}(x,x')
 &=-\frac{i}{2\pi}\,e^{im(\phi-\phi')}\\
 &\quad\times{\sum_n} e^{-i{}_s\omega_{\ell mn}(t-t')}{}_s\mathcal R_{\ell mn},
 \end{aligned}
 \label{e:sum}
\end{equation}
so that the residue ${}_s\mathcal R_{\ell mn}$ can be associated to the (constant) QNM amplitude in the Green function.  
The sum includes all QNM poles at fixed $s,\ell,m$, with
either sign of $\omega_{\rm R}$. 

For a fixed multipole $(\ell,m)$, the complete response contains both the pole sum in Eq.~\eqref{e:sum} and the rest of the frequency integral, which we will not consider in this article. To obtain the full response to a point source, we sum these multipole responses over all $(\ell,m)$ at fixed spin weight $s$, as in Eq.~\eqref{e:retardedtransform}.

At fixed source and observer positions, a more strongly damped mode decays faster in time, but its amplitude can be larger. Suppose the complete amplitudes along one sequence of poles obey the upper bound
\begin{equation}
 |\mathcal R_n|\le
 e^{-\Gamma\omega_{{\rm I},n}+o(|\omega_{{\rm I},n}|)}
 \label{e:upper}
\end{equation}
at large damping $-\omega_{\rm I}$. Here $\Gamma$ bounds the exponential growth per unit damping; a power of $-\omega_{{\rm I},n}$ can still multiply the exponential. Including the time dependence in Eq.~\eqref{e:sum} gives
\begin{equation}
 \big|e^{-i\omega_n(t-t')}\mathcal R_n\big|\le
 e^{(t-t'-\Gamma)\omega_{{\rm I},n}+o(|\omega_{{\rm I},n}|)} .
 \label{e:poletermexample}
\end{equation}
Since the asymptotic QNM sequences considered here have
$-\omega_{{\rm I},n}\sim\alpha n$, with $\alpha>0$
~\cite{keshet2008,keshetBenMeir2014}, the bound in
Eq.~\eqref{e:poletermexample} decreases exponentially
with $n$ whenever $t-t'>\Gamma$.
Comparison with a geometric series therefore gives
absolute convergence of that sequence's contribution
to Eq.~\eqref{e:sum} for
\begin{equation}
 t-t'>\Gamma .
 \label{e:sectortime}
\end{equation}
The aim of this article is to understand how $\Gamma$ depends on where we place the source and observer. In Eq.~\eqref{e:mainres}, their radii $r,r'$ enter through the two radial solutions and their angles $\theta,\theta'$ through the projector. We now examine these factors at large damping.

\section{Radial and angular growth at high damping}\label{s:wkb}
\colorlet{DCBeforeSecIII}{DCDarkOrange}
\definecolor{DCDarkOrange}{HTML}{846600}

To apply the convergence criterion $t-t'>\Gamma$, we need the amplitude growth rate at the chosen source and observer positions. The radial solutions and the angular projector in each amplitude both depend on the angular eigenvalue $A={}_s A_{\ell m}(a\omega)$. Studies of Kerr QNMs have found two behaviors of $A$ in the limit $-\omega_{\rm I}\to\infty$ with bounded $\omega_{\rm R}$~\cite{bertiAngular2006,vickersCook2022,keshetBenMeir2014}. Linear growth with $\omega$ characterizes the \textit{prolate} sector and quadratic growth the \textit{oblate} sector. For the prolate angular ground state and the oblate branch studied here, the leading terms are
\begin{equation}
 A_{\pr}=ia\omega+\mathcal{O}(1),\qquad
 A_{\ob}=-(a\omega)^2+\mathcal{O}(\omega).
 \label{e:Abalances}
\end{equation}
The radial equation contains the combination $\lambda=A+a^2\omega^2-2am\omega$. In the prolate sector $a^2\omega^2$ remains the leading term in $\lambda$ for large $\omega$. In the oblate sector the quadratic terms cancel, leaving $\lambda=O(\omega)$. The two sectors therefore have different leading radial wave numbers. 
We take $a=0.7$, $s=-2$ and $m=2$, with fixed source and observer radii $r,r'>r_+$ and polar angles $\theta,\theta'\in(0,\pi)$.

To construct the amplitudes at high damping, we first need
the radial and angular wave profiles. The basic idea comes
from the Wentzel--Kramers--Brillouin (WKB) approximation: putting the separated equations
in Schr\"odinger-type form makes their leading exponential
behavior explicit. For example, the radial solution has the schematic form
\begin{equation*}
 {}_s R^{\IN}_{\ell m\omega}(r) \sim
 c_+e^{i\omega D_\sigma(r)}
 +c_-e^{-i\omega D_\sigma(r)},
\end{equation*}
where $D_\sigma$ is the real radial action whose reference
is fixed below. Since $\omega_{\rm I}<0$, one exponential
factor grows outward and the other inward. They exchange
dominance at the radius where their magnitudes balance.
We call this the \textit{bounce radius}, characterized by
$D_\sigma(r_{\rm b}^\sigma)=0$. On either side, the larger
factor grows as $e^{(-\omega_{\rm I})|D_\sigma|}$.
This suggests a contribution $|D_\sigma|$ to $\Gamma$
at both the source and the observer.
The same goes for the angular solution, but we must also
impose regularity at the poles. To obtain the complete
amplitude growth rate, we must check the WKB coefficients,
angular normalization, Wronskian derivative and possible
cancellations. 

\subsection{Radial profiles, optical actions and growth}
Removing the first derivative puts the radial Teukolsky equation in Schrödinger-type form,
\begin{subequations}
\begin{equation}
 {}_s\widetilde R_{\ell m\omega}=\Delta^{(s+1)/2}{}_s R_{\ell m\omega},
 \qquad \partial_r^2\widetilde R+Q_r\widetilde R=0
 \label{e:rescalings}
\end{equation}
Here $Q_r$ is the square of the local radial momentum, \begin{align}  Q_r={}&\frac{K^2-2is(r-1)K}{\Delta^2}        +\frac{4is\omega r-\lambda}{\Delta}\nonumber\\  &-\frac{s+1}{2}\frac{\Delta''}{\Delta}        +\frac{1-s^2}{4}\left(\frac{\Delta'}{\Delta}\right)^2.  \label{e:Qr} \end{align} The two local WKB solutions are
\begin{equation}
 \widetilde R_\pm(r)\simeq Q_r(r)^{-1/4}
 \exp\!\left[\pm i\int^r\sqrt{Q_r(u)}\,\dd u\right]
 \label{e:localradialwkb}
\end{equation}
\end{subequations}
This local form holds away from turning points.
Defining $q_0=(r^2+a^2)^2-a^2\Delta$, the identity
\begin{equation}
 K^2-\lambda\Delta
 =\omega^2q_0-A\Delta-4am\omega r+a^2m^2
 \label{e:K}
\end{equation}
gives, at fixed $r>r_+$ and large damping~\cite{keshet2008},
\begin{equation}
 Q_r=\frac{\omega^2q_0-A\Delta}{\Delta^2}+O(\omega).
 \label{e:radialleading}
\end{equation}
For prolate, $A=O(\omega)$ leaves the $\omega^2 q_0$ term intact.
For oblate, $-A\Delta$ cancels the term $-a^2\omega^2\Delta$ in
$\omega^2 q_0$. Write $\sqrt{Q_r}=\omega D'_\sigma+O(1)$, where
$\sigma=\pr,\ob$ and a prime denotes $\dd/\dd r$. We choose the
square root for which $D'_\sigma(r)>0$ at real radii $r>r_+$. This gives
\begin{equation}
 D'_{\pr}=\frac{\sqrt{q_0}}{\Delta} = (r_* - \delta r_*)',\qquad
 D'_{\ob}=\frac{r^2+a^2}{\Delta}=r_*'
 \label{e:slopes}
\end{equation}
Here $r_*$ is the tortoise coordinate in Eq.~\eqref{eq:tortoise}.
Subtracting the two slopes and integrating from the outer horizon gives
\begin{equation}
 \delta r_*(r)=\int_{r_+}^r
 \frac{a^2\,\dd u}{u^2+a^2+\sqrt{q_0(u)}} .
 \label{eq:drs_fin}
\end{equation}
The correction increases with radius and stays finite at infinity.

The two signs in Eq.~\eqref{e:localradialwkb} give two exponential
factors in the radial solution entering the amplitude in Eq.~\eqref{e:mainres}.
To compare them at the source or observer, we measure both phases
in each sector from the same complex turning point, where the
leading radial momentum vanishes. We continue the branch with
$D'_\sigma>0$ for real $r>r_+$ into the complex $r$ plane and
integrate to the source or observer radius.
For prolate the turning point is the first-quadrant zero
$r_{\rm tp}$ of $q_0$~\cite{oshita2026}.
For oblate, the leading squared momentum has double zeros
at $\pm ia$; we use $ia$. The actions are

\begin{subequations}
\begin{align}
 z_{\pr}(r)&=\int_{r_{\rm tp}}^r D'_{\pr}(u)\,\dd u,\nonumber\\
 z_{\ob}(r)&=\int_{ia}^r D'_{\ob}(u)\,\dd u
 \label{e:complexactions}
\end{align}
Both paths pass above the horizons without winding around them.
The tortoise coordinate in Eq.~\eqref{eq:tortoise} is continued
along the integration path in the complex plane, with its
logarithms real for $r>r_+$. In particular,
$z_{\ob}(r)=r_*(r)-r_*(ia)$.

The horizon-ingoing solution $R_{\IN}={}_s R^{\IN}_{\ell m\omega}$
is a combination of the two WKB solutions. To leading exponential order,
\begin{equation}
 R_{\IN}(r)\simeq
 \widehat c_+(\omega)e^{i\omega z_\sigma(r)}
 +\widehat c_-(\omega)e^{-i\omega z_\sigma(r)}
 \label{e:turningpointwaves}
\end{equation}
\end{subequations}
At fixed radius, the WKB prefactor in Eq.~\eqref{e:localradialwkb} scales as
$|\omega|^{-1/2}$ in magnitude, so it contributes zero to the limiting growth rate $\Gamma$. 
The two exponential factors in Eq.~\eqref{e:turningpointwaves} have equal
magnitude on an anti-Stokes line, where $\Ren(i\omega z_\sigma)=0$~\cite{keshet2008}.
At the crossing of this line with the real axis outside the horizon, 

$\Ren z_\sigma=-(\omega_{\rm R}/\omega_{\rm I})\Imn z_\sigma$ tends to zero
as $-\omega_{\rm I}\to\infty$ with bounded $\omega_{\rm R}$.
This limiting crossing defines the bounce radius in each sector,
\begin{equation}
 \Ren z_{\pr}(\rbpr)=0,\qquad \Ren z_{\ob}(\rbob)=0\,
 \label{e:bounce}
\end{equation}

This locates the high-damping balance point anticipated
at the start of this section, where the two radial
exponential factors exchange dominance.

\begin{figure}[!t]
\centering
\includegraphics[width=.9 \columnwidth]{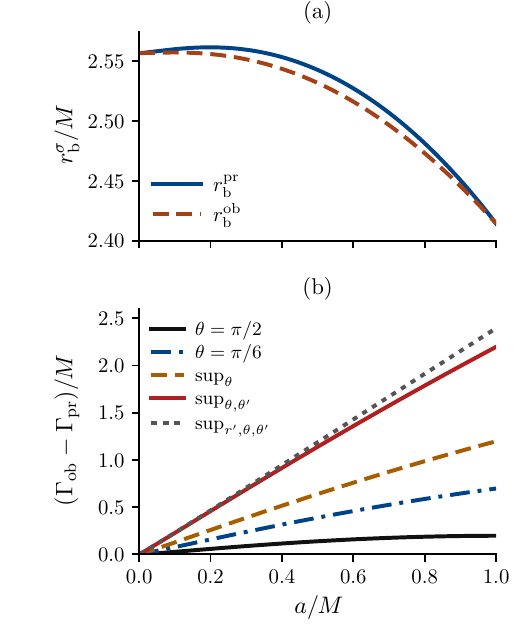}
\caption{\label{f:bouncespin}(a) Bounce radii as functions
of spin.
(b) Difference between the convergence thresholds for an
observer at infinity,
$\lim_{r\to\infty}(\Gamma_{\ob}-\Gamma_{\pr})/M$.
Black and blue: source at $r'=\rbpr$, $\theta'=\pi/2$,
with observer angles $\theta=\pi/2$ and $\pi/6$, respectively.
Ochre: supremum over $\theta$ with the same source.
Red: supremum over both angles at $r'=\rbpr$.
Gray dotted: supremum over both angles and the source
radius $r'>r_+$.
See Sec.~\ref{s:competition} for discussion.}
\end{figure}

We find $\rbob<\rbpr$ for $0<a<M$, as shown in Fig.~\ref{f:bouncespin}.
The two radii are close: the largest difference is
about $0.01M$ near $a/M\simeq0.6$. At $a/M=0.7$ both are close to $2.5M$.
For real $r>r_+$, the slopes $D'_\sigma$ are real and nonzero,
so $\Imn z_\sigma$ is constant there. The real part is the signed \textit{optical distance} from the bounce,
\begin{equation}
 D_\sigma(r)=\int_{r_{\rm b}^{\sigma}}^r D'_\sigma(u)\,\dd u
            =\Ren z_\sigma(r)
 \label{e:actions}
\end{equation}

Using Eq.~\eqref{e:slopes} and the reference radius of each sector gives
\begin{subequations}
    \begin{align}
        D_{\ob}(r) &= r_*(r)-r_*(\rbob) \label{eq:Dob} \; , \\
        D_{\pr}(r) &= r_*(r) - \delta r_*(r) - r_*(\rbpr) + \delta r_*(\rbpr) \label{eq:Dpr} \; 
    \end{align}
\end{subequations}

Writing $z_\sigma(r)=D_\sigma(r)+i\beta_\sigma$ with
$\beta_\sigma=\Imn z_\sigma(r_{\rm b}^\sigma)$ gives
$e^{\pm i\omega z_\sigma}=e^{\pm i\omega D_\sigma}e^{\mp\omega\beta_\sigma}$.
The second factor has a bounded magnitude $e^{\mp\omega_{\rm R}\beta_\sigma}$
and an oscillating phase $\mp\omega_{\rm I}\beta_\sigma$. For the radial growth check performed in this section we absorb
it into $c_\pm=\widehat c_\pm e^{\mp\omega\beta_\sigma}$, but the phase comparison
in App.~\ref{s:complex} retains the full action $z_\sigma$. The radial solution can therefore be written as
\begin{equation}
 R_{\IN}(r)\simeq c_+(\omega)e^{i\omega D_\sigma(r)}
                   +c_-(\omega)e^{-i\omega D_\sigma(r)}
 \label{e:waves}
\end{equation}

Above $r_{\rm b}^\sigma$, $D_\sigma>0$ and the plus exponential grows.
Below $r_{\rm b}^\sigma$, $D_\sigma<0$ and the minus exponential grows.
The larger of the two exponential factors has rate
\begin{equation}
 \max\{D_\sigma(r),-D_\sigma(r)\}=|D_\sigma(r)|
 \label{e:carriergrowth}
\end{equation}
The rescaling between $R$ and $\widetilde R$ at fixed radius does not change this rate. This recovers the intuition outlined at the start of
this section: the two exponential factors balance at the
bounce, and the larger grows with the optical distance
$|D_\sigma|$ on either side.

In order to numerically check the scaling of $R_\IN$ at large $-\omega_{\rm I}$ in Eq.~\eqref{e:waves}, we plot the ratio of $R_\IN$ divided by its value at the bounce, 
\begin{equation}
 \frac{\log|R_{\IN,n}(r)/R_{\IN,n}(r_{\rm b}^{\sigma})|}
      {-\omega_{{\rm I},n}}\longrightarrow |D_\sigma(r)|
 \label{e:radialratemain}
\end{equation}
This choice of normalization cancels any common exponential term in $c_\pm$ in Eq.~\eqref{e:waves}; additionally, since $D_\sigma(r_{\rm b}^\sigma)=0$, the predicted limiting rate at large damping of this ratio is simply $|D_\sigma(r)|$. 

\begin{figure*}[!t]
\centering\includegraphics[width=.9\textwidth]{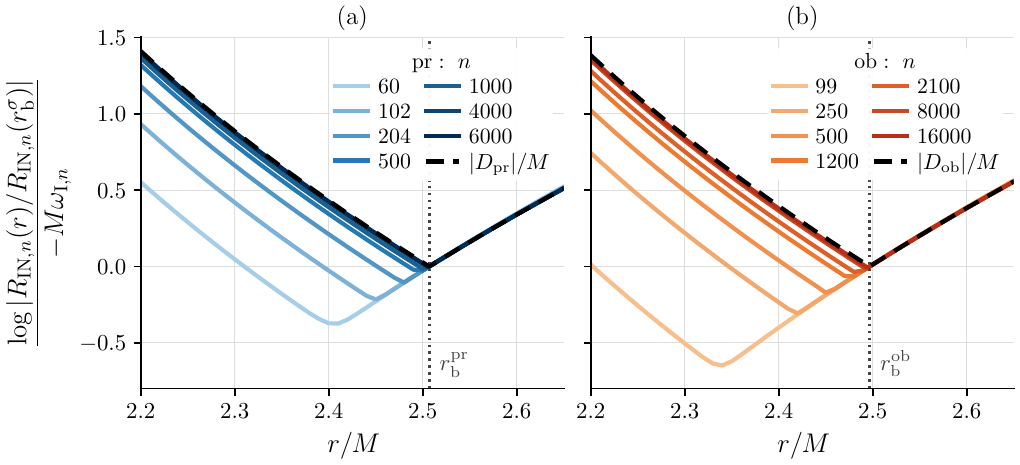}
\caption{\label{f:radialprofiles}Radial profiles at $a/M=0.7$,
normalized at the corresponding bounce radius.
Each numerical curve shows
$\log|R_{\IN,n}(r)/R_{\IN,n}(r_{\rm b}^{\sigma})|/(-M\omega_{{\rm I},n})$.
The legends give $n$.
Dashed curves show $|D_\sigma(r)|/M$; vertical dotted lines
mark the bounce radii.
See Appendix~\ref{a:numerical} for the numerical method.}
\end{figure*}

Figure~\ref{f:radialprofiles} tests this prediction on both sides of the bounce.
Near the bounce the logarithm can be negative at finite damping; at the bounce itself it is zero by normalization.

\Needspace{5\baselineskip}

\subsection{Angular profiles and growth}

At fixed radii the projector in Eq.~\eqref{eq:angularProj} determines how
the amplitude changes with the source and observer angles. As in the radial problem, we obtain two local
exponential solutions. The angular profile must also
be regular at both poles. To determine the allowed
combination, we must connect these approximations to
solutions valid near the poles and, for prolate,
near $\theta=\pi/2$. We first remove the first derivative from Eq.~\eqref{e:TeukS},
\begin{subequations}
\begin{equation}
 \widetilde S=\sqrt{\sin\theta}\,S,\qquad
 \partial_\theta^2\widetilde S+Q_\theta\widetilde S=0,
 \label{e:angularrescaling}
\end{equation}
where
\begin{align}
 Q_\theta={}&(a\omega)^2\cos^2\theta-2a\omega s\cos\theta
                 +A+s+\tfrac14\nonumber\\
 &-\frac{(m+s\cos\theta)^2-\tfrac14}{\sin^2\theta}.
 \label{e:Qtheta}
\end{align}
\end{subequations}
The WKB exponents are
$\pm i\int^\theta\sqrt{Q_\theta(\vartheta)}\,\dd\vartheta$.
For large $-\omega_{\rm I}$ with bounded $\omega_{\rm R}$,
Eq.~\eqref{e:Abalances} gives
\begin{subequations}
\begin{equation}
 Q_\theta\simeq
 \begin{cases}
 -(a\omega_{\rm I})^2\cos^2\theta,&\pr,\\
  (a\omega_{\rm I})^2\sin^2\theta,&\ob.
 \end{cases}
 \label{e:angularmomenta}
\end{equation}
At leading order, these signs give exponential behavior for prolate and oscillatory behavior for oblate. As in the original angular eigenvalue problem, we require $S$ to
be regular at $\theta=0,\pi$. To construct the angular profile between them,
we must also treat the regions where WKB fails, i.e. turning points and singular endpoints: near the poles
in both sectors, and near the equator for prolate. Indeed, the prolate
leading potential in Eq.~\eqref{e:angularmomenta} vanishes at
$\theta=\pi/2$, while the oblate one vanishes at $\theta=0,\pi$.

For prolate, integrating from $\theta=\pi/2$ gives
$\int_{\pi/2}^{\theta}\cos\vartheta\,\dd\vartheta=\sin\theta-1$,
and hence
\begin{equation}
 \widetilde S\propto e^{+ia\omega(1-\sin\theta)}
 \quad\text{or}\quad e^{-ia\omega(1-\sin\theta)}.
 \label{e:prcarrier}
\end{equation}
We first impose regularity at $\theta=0$. For the prolate branch
$A_{\pr}=ia\omega+O(1)$, the regular local solution grows in magnitude
as we move towards the equator, in the region where it overlaps
with WKB (Appendix~\ref{a:angular}). Matching therefore selects the
minus sign in Eq.~\eqref{e:prcarrier}. Regularity at $\theta=\pi$
selects the same sign on the other side of the equator.

The two WKB expressions are connected through the
equatorial region by a Gaussian local solution
(Appendix~\ref{a:angular}).

At a fixed angle away from the poles, $\sqrt{\sin\theta}$
is independent of frequency. With the normalization fixed by
the equatorial solution, the remaining prefactor adds no
exponential rate. Taking the modulus of the minus
branch in Eq.~\eqref{e:prcarrier}, we obtain
\begin{equation}
 \log\left|\frac{S(\theta)}{S(\pi/2)}\right|
 =a\omega_{\rm I}(1-\sin\theta)+o(|\omega_{\rm I}|).
 \label{e:angcarrier}
\end{equation}
\end{subequations}

For oblate the leading potential has no zero between the poles.
We can impose regularity and select the WKB sign in the same
local calculation near $\theta=0$. Integrating the leading
momentum from that pole gives
\begin{equation}
 \widetilde S\propto e^{+a\omega(1-\cos\theta)}
 \quad\text{or}\quad e^{-a\omega(1-\cos\theta)}.
 \label{e:obcarrier}
\end{equation}
For $s=-2$, $m=2$ and $A_{\ob}=-(a\omega)^2-2a\omega+O(1)$,
set $v=a\omega(1-\cos\theta)$. At fixed $v$ the leading
local equation and its regular solution are~\cite{casalsOttewill2005}
\begin{equation}
 \partial_v(v \partial_v S)+(1-v)S=0,\qquad S=e^{-v}.
 \label{e:obpolar}
\end{equation}
The other linearly independent solution is logarithmically singular at $v=0$, so we discard it.
Matching $e^{-v}$ to Eq.~\eqref{e:obcarrier} selects the minus
sign. Both WKB exponential factors have bounded magnitude
because $\omega_{\rm R}$ stays bounded, but this sign matters
for the phase test in Appendix~\ref{s:complex}.

Finally, the projector
$\Pi_n=S_n(\theta)S_n(\theta')/N_n$
contains one angular profile evaluated at the observer
and one at the source. Using the norm estimates in
Appendix~\ref{a:angular}, division by $N_n$ adds only a power
of $-\omega_{\rm I}$. The exponential rate of the projector
is therefore the sum of the rates of these two factors.
At fixed angles away from $\theta=0,\pi$,
\begin{equation}
 \lim_{n\to\infty}\frac{\log|\Pi_n|}{-\omega_{{\rm I},n}}
 =
 \begin{cases}
 -a(2-\sin\theta-\sin\theta'),&\pr,\\
 0,&\ob.
 \end{cases}
 \label{e:angularrates}
\end{equation}
For the equatorial source in Fig.~\ref{f:angularprofiles}, these
predictions reduce to $-a(1-\sin\theta)$ for prolate and zero for oblate.

\onecolumngrid
\noindent\begin{minipage}{\textwidth}
\makeatletter\def\@captype{figure}\makeatother
\kern 3pt
\centering\includegraphics[width=.9\textwidth]{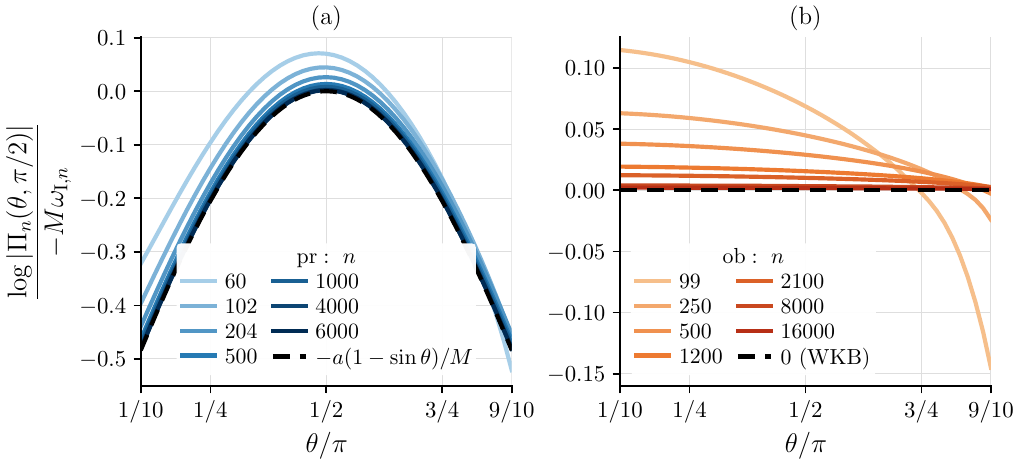}
\caption{\label{f:angularprofiles}Angular projector at $a/M=0.7$,
with the source on the equator.
Solid curves show the numerical values of
$\log|\Pi_n(\theta,\pi/2)|/(-M\omega_{{\rm I},n})$.
The legends give $n$.
Dashed lines show the WKB predictions: $-a(1-\sin\theta)/M$
for prolate and zero for oblate.
Near $\theta=\pi$, the WKB profile requires a separate polar
approximation, discussed in Appendix~\ref{a:angular}.
The panels have different vertical scales.}

\end{minipage}\par\smallskip
\twocolumngrid

\section{Oblate dominance and the equatorial prolate region}\label{s:competition}

Section~\ref{s:wkb} gives the radial and angular profiles and
tests them numerically. We now combine these results into the growth rate
of each complete amplitude and compare the two sectors.
The larger rate sets the later convergence threshold,
since more time is needed for temporal decay to overcome
amplitude growth. The complete amplitude in
Eq.~\eqref{e:mainres} also contains the Wronskian derivative,
which is independent of the source and observer positions
but changes from one mode to the next. To separate the spatial
profiles from the remaining radial factor, we use the same
reference radius as before and define the radial QNM amplitude
with both radii at the bounce,
\begin{equation}
 \mathcal R_n^{\rm rad}(r_{\rm b}^\sigma,r_{\rm b}^\sigma)
 =
 \left.
 \frac{{}_s R^{\IN}_{\ell m\omega}(r_{\rm b}^\sigma)
       {}_s R^{\UP}_{\ell m\omega}(r_{\rm b}^\sigma)}
      {\dd{}_s\widehat W_{\ell m}/\dd\omega}
 \right|_{\omega=\omega_n}.
 \label{e:Rb}
\end{equation}

The two radial solutions are proportional at a QNM frequency.
Hence, provided $R_{\IN,n}(r_{\rm b}^\sigma)\ne0$,
Eq.~\eqref{e:mainres} can be rewritten exactly as
\begin{equation}
 \mathcal R_n=\Pi_n(\theta,\theta')\,
 \mathcal R_n^{\rm rad}(r_{\rm b}^\sigma,r_{\rm b}^\sigma)
 \frac{R_{\IN,n}(r)}{R_{\IN,n}(r_{\rm b}^\sigma)}
 \frac{R_{\IN,n}(r')}{R_{\IN,n}(r_{\rm b}^\sigma)}.
 \label{e:product}
\end{equation}
Each radial ratio contributes $|D_\sigma|$ at its corresponding
radius [Eq.~\eqref{e:radialratemain}], while the projector supplies
the angular rate in Eq.~\eqref{e:angularrates}.
Multiplying these factors adds their exponents:\begin{equation}
 |\mathcal R_{\sigma,n}|
 =
 |\mathcal R^{\rm rad}_{\sigma,n}
 (r_{\rm b}^{\sigma},r_{\rm b}^{\sigma})|\,
 e^{-\omega_{{\rm I},n}\Gamma_\sigma
    +o(|\omega_{{\rm I},n}|)},
 \label{e:spatialexponentials}
\end{equation}
where
\begin{subequations}
\begin{align}
 \Gamma_{\pr}
 &=|D_{\pr}(r)|+|D_{\pr}(r')|
   -a(2-\sin\theta-\sin\theta'),
 \label{e:gpr}\\
 \Gamma_{\ob}
 &=|D_{\ob}(r)|+|D_{\ob}(r')|.
 \label{e:gob}
\end{align}
\end{subequations}
We assume that the reference amplitude in Eq.~\eqref{e:Rb}
adds no further exponential rate. This assumption is numerically verified in Appendix~\ref{a:numerical}.
 To compare the two rates at $a/M=0.7$, let $J(r)=|D_{\ob}(r)|-|D_{\pr}(r)|$ be the difference of the two optical distances at one radius. Then
\begin{align}
 \Gamma_{\ob}-\Gamma_{\pr}
 &=J(r)+J(r')\nonumber\\
 &\quad+a(2-\sin\theta-\sin\theta') \label{e:comparison}
\end{align}
Oblate has the larger
amplitude growth coefficient if this difference is positive,
and prolate if it is negative. 
The angular term vanishes when both endpoints lie on the equator and is positive whenever either endpoint is off the equator.

\begin{figure}[!t]
\centering
\includegraphics[width=\columnwidth]{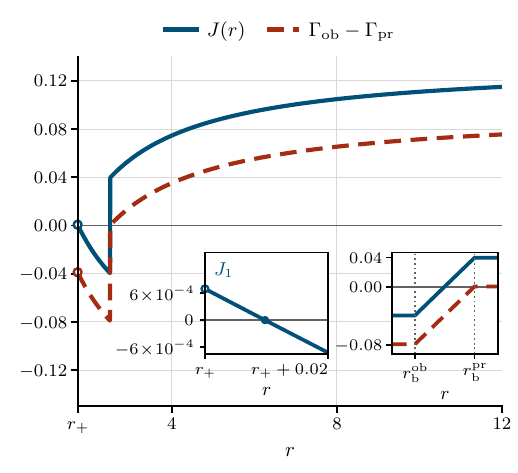}
\caption{\label{f:actions}Optical-distance difference $J(r)$
(solid) and rate difference $\Gamma_{\ob}-\Gamma_{\pr}$
(dashed), with $M=1$, $a=0.7$, $r'=\rbob$ and
$\theta=\theta'=\pi/2$.
The left inset resolves $J(r)$ near the horizon and its zero
crossing; the right inset enlarges both curves near the bounce
radii, marked by vertical dotted lines.}
\end{figure}

On the other hand, using Eqs.~\eqref{eq:Dob}-\eqref{eq:Dpr}, the fact that $D_\sigma$ changes sign at its corresponding bounce radius, and the inequality $\rbob < \rbpr$, we find
\begin{equation}
 J(r)=
 \begin{cases}
 J_1 - \delta r_*(r) ,&r_+<r\le\rbob\\
 2 r_*(r) - \delta r_*(r) - J_2,&\rbob<r<\rbpr\\
 \delta r_*(r) - J_1,&r\ge\rbpr
 \end{cases}
 \label{e:piece}
\end{equation}
where $J_1 = r_*(\rbob) - r_*(\rbpr) + \delta r_*(\rbpr)$ and $J_2 = r_*(\rbob) + r_*(\rbpr) - \delta r_*(\rbpr)$. A plot of $J(r)$ is shown in Fig.~\ref{f:actions}; its minimum is $J(\rbob)=-|D_{\pr}(\rbob)|<0$. The dashed curve gives the smallest possible $\Gamma_{\ob}-\Gamma_{\pr}$ at each observer radius, obtained for $r'=\rbob$ and $\theta=\theta'=\pi/2$. Wherever it is positive, oblate dominates for any source radius and angles.  If any of the endpoints $r,r'$ reaches or exceeds $\rbpr$, then  $J(r)+J(r') \ge J(\rbpr)+J(\rbob) > 0$. Thus
\begin{equation}
 \max(r,r')\ge\rbpr\quad\Longrightarrow\quad
 \Gamma_{\ob}>\Gamma_{\pr}
 \label{e:strict}
\end{equation}

\begin{figure*}[t]
\centering\includegraphics[width=.98\textwidth]{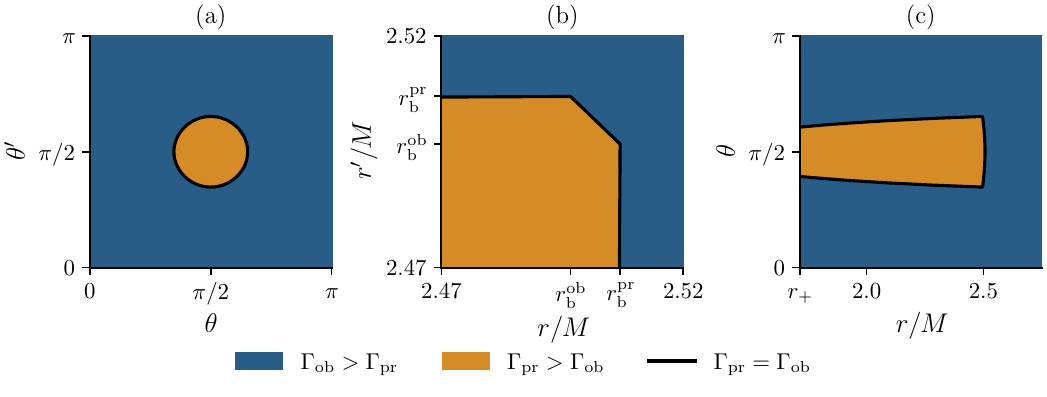}
\caption{\label{f:atlas}Plot of the region of dominance for each WKB rate at $a/M=0.7$: in blue the oblate rate dominates, in orange the prolate one dominates. Curves mark equal rates. (a) Dependence on both angles at fixed $r=r'=\rbob$.
(b) Dependence on both radii at fixed $\theta=\theta'=\pi/2$.
(c) Dependence on the observer position for a source at $(\rbob,\pi/2)$.
The rate formulas are plotted up to the horizon and the poles;
the WKB approximations are not uniform near these boundaries.  }
\end{figure*}

For both radii below $\rbpr$, prolate can dominate for some
choice of angles if and only if $J(r)+J(r')<0$.
For fixed radii $r,r'$, the difference in Eq.~\eqref{e:comparison} is negative precisely when
\begin{equation}
 \sin\theta+\sin\theta'>
 2+\frac{J(r)+J(r')}{a}
 \label{e:prolateangles}
\end{equation}
The most favorable radii for prolate are $r=r'=\rbob$,
where both radial contributions attain their minimum.
At $a/M=0.7$, the lowest angular threshold is therefore
\begin{equation}
 2+\frac{2J(\rbob)}{a}\simeq1.89.
 \label{e:angularthreshold}
\end{equation}
Since neither sine exceeds one, both the source and the
observer must lie near the equator for prolate to dominate.
Figure~\ref{f:atlas} shows the region of dominance of each sector for different observer and source positions.

Using the asymptotic description above and the damping
condition stated in Sec.~\ref{s:method}, both pole sums converge absolutely for
\begin{equation}
 t-t'>\max(\Gamma_{\pr},\Gamma_{\ob}) .
 \label{e:sufficient}
\end{equation}
For $t-t'<\Gamma_\sigma$, the terms of the corresponding
sum grow exponentially, so the sum diverges. 
Between the two thresholds, the sector with the smaller
$\Gamma_\sigma$ therefore converges while the other diverges.
The larger rate determines when both sums converge.
Notice that the close proximity of the two bounce radii
does not imply that the two families of modes converge
at nearly the same time in the Green function.

Indeed, prolate and oblate modes 
have different radial propagation speeds
(cf. Eq.~\eqref{e:slopes}) so they can accumulate a larger
time difference. For definiteness, let us consider an
observer at very large distance from the BH and a source
at the prolate bounce $\rbpr$, both of them in the
equatorial plane. The black curve in the 
lower panel of Fig.~\ref{f:bouncespin} plots the difference
between the convergence times of the two families of modes,
which depends only on the spin of the BH.
It can reach $0.2M$ for the extremal case, which is much larger than the difference in light-travel time
between the two bounce radii. 
The other curves show how this gap increases when
the angular positions and the source radius are allowed
to vary. The angular contribution in
Eq.~\eqref{e:comparison} can add up to $2a$. 
Allowing both angles and the source radius to vary gives
the gray curve, which approaches about $2.4M$ for
$r'\to\infty$, $\theta,\theta'\to0$ and $a/M\to1$.

Finally, let us remark that the time entering the criterion~\eqref{e:sufficient} need not be equal to the first causal arrival between the two positions. For different polar angles, any causal curve satisfies $t-t'\ge r_+|\theta-\theta'|$, independently of azimuthal separation (Appendix~\ref{a:optical_geometry}). At $a/M=0.7$, taking $(r',\theta')=(2.4M,\pi/2)$ and $(r,\theta)=(2.6M,\pi/6)$, the oblate rate sets the criterion at $0.78M$ while the causal lower bound is $1.8M$. At $t-t'=M$, the two pole sums at fixed $(\ell,m)$ therefore converge, although the full retarded response to the impulse is still zero. Notice that causal support concerns the complete response, including the sum over angular multipoles and the remaining frequency-integral contributions.

\section{Bounce surfaces and the geometry of QNM convergence}\label{s:geometry}
In this Section, we show that the source and observer contributions
to $\Gamma_\sigma$ also admit a geometric interpretation.
The idea is that spatial growth competes with temporal
decay. For a fixed source, this competition is:
\begin{equation*}
 e^{-i\omega(t-t')}e^{i\omega\Gamma_\sigma(r,\theta)}
 =e^{-i\omega[(t-t')-\Gamma_\sigma(r,\theta)]}.
\end{equation*}
This suggests interpreting $t'+\Gamma_\sigma(r,\theta)$
as the passage time of a front. For a light front, the
surface $t=t'+\Gamma_\sigma(r,\theta)$ must be null.
We first examine the radial actions using light rays,
then test whether the angular dependence admits this
interpretation. We finally interpret the separate source
and observer contributions in terms of their bounce surfaces.

\subsection{From radial light rays to null fronts}

To interpret the radial actions $D_{\pr}$ and $D_{\ob}$,
we compare their derivatives with the coordinate travel time
of null geodesics. Consider a photon with energy $E=-p_t>0$
and axial angular momentum $L_z=0$. 
Carter's equations for $t,r,\theta$ are~\cite{carter1968,viFerrariGualtieriPani}
\begin{equation}
 \begin{aligned}
 \Sigma\dot t&=E\left[\frac{(r^2+a^2)^2}{\Delta}
                         -a^2\sin^2\theta\right],\\
 (\Sigma\dot r)^2&=E^2(r^2+a^2)^2-\Delta(Q+a^2E^2),\\
 (\Sigma\dot\theta)^2&=Q+a^2E^2\cos^2\theta.
 \end{aligned}
 \label{e:carter}
\end{equation}
Here $Q$ is the Carter constant,
$\Sigma=r^2+a^2\cos^2\theta$, and dots denote affine-parameter
derivatives. For $Q=0$, the polar potential $Q+a^2E^2\cos^2\theta$ and its angular
derivative vanish at $\theta=\pi/2$, allowing a photon to remain
on the equator. For $Q=-a^2E^2$, they vanish at $\theta=0,\pi$,
allowing motion along the rotation axis.

The time per unit radial interval is $\dot t/|\dot r|$.
Substituting these two choices gives
\begin{equation}
 \begin{aligned}
 \left.\frac{\dot t}{|\dot r|}\right|_{\rm eq}
 &=\frac{E q_0/\Delta}{E\sqrt{q_0}}
 =\frac{\sqrt{q_0}}{\Delta}=D'_{\pr}(r),\\
 \left.\frac{\dot t}{|\dot r|}\right|_{\rm axis}
 &=\frac{E(r^2+a^2)^2/\Delta}{E(r^2+a^2)}
 =\frac{r^2+a^2}{\Delta}=D'_{\ob}(r),
 \end{aligned}
 \label{e:rays}
\end{equation}
These two rays reproduce the WKB radial slopes in
Eq.~\eqref{e:slopes}. To make the connection explicit, we
write Eq.~\eqref{e:radialleading} and the radial Carter
equation in the same form:
\begin{equation}
 \begin{aligned}
 \frac{\Delta^2Q_r}{\omega^2}
 &=q_0-\Delta\frac{A}{\omega^2}+O(\omega^{-1}),\\
 \left(\frac{\Sigma\dot r}{E}\right)^2
 &=q_0-\Delta\frac{Q}{E^2}.
 \end{aligned}
 \label{e:carter_wkb}
\end{equation}
Thus, at fixed $m$ and large damping, $A/\omega^2$ plays
the role of $Q/E^2$~\cite{keshetBenMeir2014}.
Equation~\eqref{e:Abalances} gives
$A_{\pr}/\omega^2\to0$ and $A_{\ob}/\omega^2\to-a^2$.
These limits reproduce the choices $Q=0$ and $Q=-a^2E^2$
for the equatorial and axial rays, respectively.

Integrating these slopes gives the radial travel time. The bounce condition fixes the reference radius through $D_\sigma(r_{\rm b}^\sigma)=0$ [Eq.~\eqref{e:bounce}], so
$D_\sigma(r)=\int_{r_{\rm b}^\sigma}^{r}D'_\sigma(u)\,\dd u$.
Since both slopes are positive outside the horizon, $D_\sigma$ is negative below its bounce and positive above it. Thus $|D_\sigma(r)|$ is the travel time between $r_{\rm b}^\sigma$ and $r$ along the corresponding special ray. 

These rays explain the radial dependence, but the amplitude also depends on the polar angles of the source and observer. We therefore ask whether the same radial slopes can be completed by an angular dependence to describe a null front. We seek an axisymmetric front $t=T(r,\theta)$ whose radial
derivative reproduces the leading WKB radial slope at large damping.
Here $T$ gives the front's passage time at each position. 
 Requiring the normal $\partial_\mu(t-T)$ to be null gives~\cite{viPretoriusIsrael}
\begin{equation}
 \Delta(\partial_rT)^2+(\partial_\theta T)^2
 =\frac{(r^2+a^2)^2}{\Delta}-a^2\sin^2\theta.
 \label{e:vi_front_eikonal}
\end{equation}
Keeping $\partial_rT=\pm D'_\sigma$, we solve for the angular part. For prolate, the radial term is $q_0/\Delta=(r^2+a^2)^2/\Delta-a^2$; for oblate, it is $(r^2+a^2)^2/\Delta$. Subtraction gives
\begin{equation}
 (\partial_\theta T)^2=
 \begin{cases}
 a^2\cos^2\theta,&\pr,\\
 -a^2\sin^2\theta,&\ob.
 \end{cases}
 \label{e:vi_angular}
\end{equation}
At nonzero spin, only the prolate slope admits a real angular completion away from the axis. We construct it first and return to oblate below.

Integrating from the equator allows either angular shift, $+a(1-\sin\theta)$ or $-a(1-\sin\theta)$: both satisfy the null condition. To select the front associated with our prolate angular profile, we use the attenuation away from the equator found in Sec.~\ref{s:wkb} [Eqs.~\eqref{e:prcarrier} and~\eqref{e:angcarrier}]. Since $|e^{i\omega T}|=e^{-\omega_{\rm I}T}$ with $-\omega_{\rm I}>0$, this attenuation requires $T(r,\theta)<T(r,\pi/2)$. The negative angular shift applies to both radial directions:
\begin{equation}
 T_\pm(r,\theta)=\pm D_{\pr}(r)-a(1-\sin\theta).
 \label{e:vi_snap}
\end{equation}
These are real Kerr null fronts on their smooth domains: the plus branch is outgoing and the minus branch ingoing.

\subsection{From the bounce radius to bounce surfaces}

We now use these fronts to interpret the source and observer
contributions to $\Gamma_{\pr}$. We write Eq.~\eqref{e:gpr}
as the sum of these two contributions and denote each by
$\tau_{\pr}$:
\begin{equation}
 \begin{aligned}
 \Gamma_{\pr}={}&
 \underbrace{|D_{\pr}(r')|-a(1-\sin\theta')}
 _{\tau_{\pr}(r',\theta')\;\text{(source)}}\\
 &+\underbrace{|D_{\pr}(r)|-a(1-\sin\theta)}
 _{\tau_{\pr}(r,\theta)\;\text{(observer)}}.
 \end{aligned}
 \label{e:vi_taupr}
\end{equation}
Below $\rbpr$, $D_{\pr}<0$, so $\tau_{\pr}=T_-$;
above $\rbpr$, $D_{\pr}>0$, so $\tau_{\pr}=T_+$,
with $T_\pm$ given in Eq.~\eqref{e:vi_snap}.

\begin{figure*}[t]
\centering
\includegraphics[width=.8\textwidth]{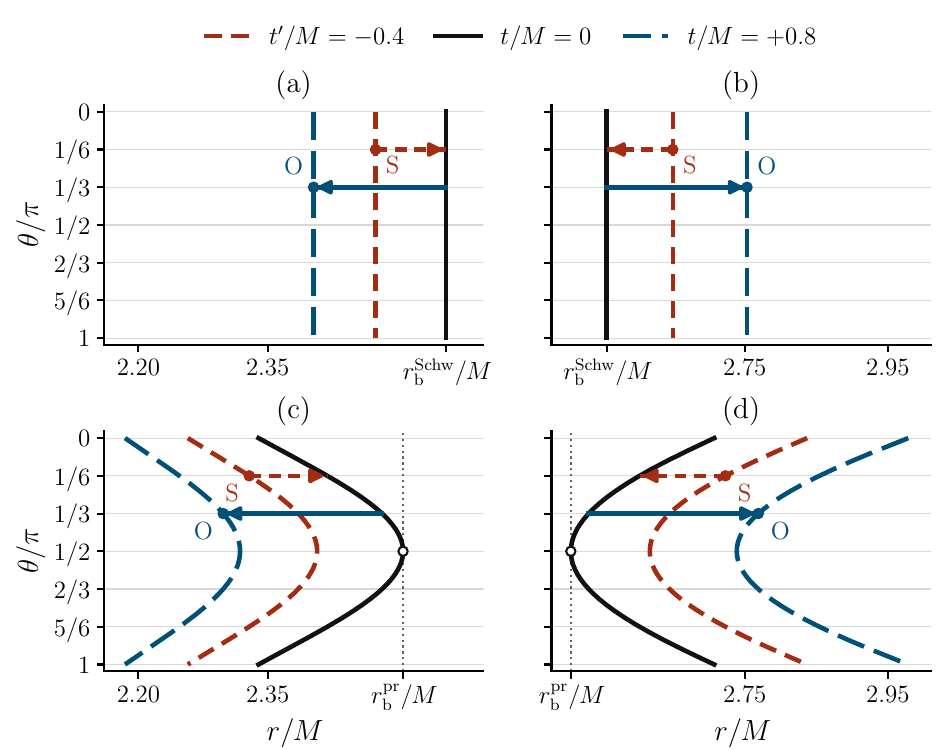}
\caption{\label{f:vi_snap}Source profiles (red), observer profiles (blue), and bounce surfaces
(black) illustrating the bounce condition for convergence of the QNM sums. Left: both positions below the radial bounce; right: both above.
Top: Schwarzschild spheres, shown as vertical lines.
Bottom: prolate Kerr at $a/M=0.7$.
The source profile is shown in reverse time order.
Black curves correspond to $t=0$.
Arrows follow the profiles at fixed polar angle, not photon paths.
The full polar range is shown, with north at the top; the WKB
approximations apply away from the poles at fixed angles.
In all four panels, $\theta'=\pi/6$ and $\theta=\pi/3$.
In the Kerr panels (c,d), oblate sets the combined convergence
criterion at these angles. }
\end{figure*}

Combining the Green function's time dependence with the
leading factor from the real optical actions gives
\begin{equation*}
 \begin{aligned}
 e^{-i\omega(t-t')}e^{i\omega\Gamma_{\pr}}
 ={}&e^{-i\omega[t-\tau_{\pr}(r,\theta)]}\\
 &e^{i\omega[t'+\tau_{\pr}(r',\theta')]}
 \end{aligned}
\end{equation*}
On a wave front, we represent the observer profile by $t=\tau_{\pr}$ and
read the source profile with the opposite time sign:
\begin{equation}
 \begin{aligned}
 t'&=-\tau_{\pr}(r',\theta'),\\
 t&=\tau_{\pr}(r,\theta).
 \end{aligned}
 \label{e:vi_endpoint_phase}
\end{equation}
Since $t=T_\pm$ gives the front's passage time, the surface
$\tau_{\pr}=0$ is reached at time $t=0$. We call it the
\textit{bounce surface} of the respective radial branch (black curves in Fig.~\ref{f:vi_snap}).

The shape of each surface follows directly from its definition:
\begin{equation}
 \tau_{\pr}=0
 \quad\Longleftrightarrow\quad
 D_{\pr}(r)=\pm a(1-\sin\theta).
 \label{e:vi_bouncesurfaces}
\end{equation}
Since $D_{\pr}$ increases from zero at $\rbpr$, the plus
sign gives the outer bounce surface, at $r\ge\rbpr$,
and the minus sign the inner one, at $r\le\rbpr$.

The radii of the two bounce surfaces depend on $\theta$ but
not on $\phi$. They meet along the equatorial circle
$r=\rbpr$, $\theta=\pi/2$. In Schwarzschild the angular
shift vanishes, and both reduce to the same sphere $D_0(r)=0$
(Appendix~\ref{a:schwarzschild}). A single bounce radius then
describes the whole surface; in Kerr its angular shape also
matters. The arrows in Fig.~\ref{f:vi_snap} follow the profiles
at fixed $\theta$, while individual light rays generally also
change polar angle (Appendix~\ref{a:optical_geometry}).

For $\tau_{\pr}(r',\theta')>0$ and $\tau_{\pr}(r,\theta)>0$,
the source profile is at $t'<0$ and the observer
profile at $t>0$, with $t-t'=\Gamma_{\pr}$.
If both positions are above $\rbpr$ (right panels), the
source profile approaches the outer bounce surface towards
smaller radii, and the observer profile departs towards
larger radii. Below $\rbpr$ (left panels), the radial
directions reverse and the reference is the inner bounce
surface.

Finally, consider the mixed case $r'<\rbpr<r$.
The source uses $T_-(r',\theta')$ and the observer
$T_+(r,\theta)$, so
\begin{equation*}
 \Gamma_{\pr}=D_{\pr}(r)-D_{\pr}(r')
 -a(2-\sin\theta-\sin\theta').
\end{equation*}
The two contributions now refer to different bounce surfaces,
except on the equator. Exchanging source and observer gives
the other mixed case.
In Schwarzschild, for positions on opposite sides of its
bounce sphere, the absolute values similarly give
$\Gamma_0=D_0(r)-D_0(r')$. At the same angular position,
this is the direct radial travel time through the bounce. In Kerr, the prolate rate contains an
additional angular subtraction. Even at $\theta'=\theta$,
it is smaller than the radial contribution
$D_{\pr}(r)-D_{\pr}(r')$ by $2a(1-\sin\theta)$, which
vanishes only on the equator at nonzero spin. This is a
reduction of the convergence threshold, not a shorter
travel time for a direct light ray.

\subsection{Oblate geometry and the convergence boundary}

For oblate at nonzero spin, Eq.~\eqref{e:vi_angular} admits no real null front with the prescribed radial slope away from the axis. The oblate angular exponential in Eq.~\eqref{e:obcarrier} corresponds to the imaginary solution $\partial_\theta T=ia\sin\theta$: it changes the phase, not the growth rate. The rate in Eq.~\eqref{e:gob} therefore contains only radial contributions, and we set $\tau_{\ob}(r)=|D_{\ob}(r)|$. Its zero set, the surface of constant Boyer--Lindquist radius $r=\rbob$, defines the oblate bounce surface. 

The oblate boundary $t-t'=\Gamma_{\ob}$ is hence
\begin{equation} \label{eq:oblateConvergence}
     t=t'+|D_{\ob}(r')|+|D_{\ob}(r)|.
\end{equation}
At any fixed time, its spatial sections have constant radius. The normal $n_\mu=\partial_\mu[t-|D_{\ob}(r)|]$ satisfies
\[
 \begin{aligned}
 n_\mu n^\mu
 &=g^{tt}+g^{rr}(D'_{\ob})^2\\
 &=\frac{a^2\sin^2\theta}{\Sigma},
 \end{aligned}
\]
using the metric in Eq.~\eqref{e:app_squaremetric}. The positive norm means that the boundary is timelike off the axis at nonzero spin: locally, its motion in the normal direction is slower than light. The norm vanishes in Schwarzschild and on the Kerr rotation axis. The oblate construction therefore retains a bounce surface without a real null wavefront away from the axis.

On the other hand, in the oblate sector the radial scattering picture generalizing the Schwarzschild case is straightforward: the time needed for the oblate QNMs to converge, Eq.~\eqref{eq:oblateConvergence}, is just the time taken by a light-ray with angular momentum $L_z=0$ and Carter constant $Q=-a^2 E^2$ to propagate on the axis $\theta=0$ from the source to the bounce radius plus the time from the bounce to the observer radius.

\section{Conclusions}\label{s:conclusions}
We have studied the QNM expansion of the retarded Green function of a Kerr black hole, i.e., the signal received at time $t$ by an observer at $(r,\theta)$ produced by an impulse emitted at time $t'$ by a source at $(r',\theta')$, focusing on the overtone sums at fixed angular indices $(\ell,m)$. Each overtone contributes a term $\mathcal R_n e^{-i\omega_n(t-t')}$ to the Green function. Its amplitude $\mathcal R_n$ depends on the two positions and can grow with the damping $-\Imn\omega_n$, while the time dependence decays with $t$. The sum, therefore, converges only once the elapsed time $t-t'$ is large enough. Our main result is that, at large damping, $\log|\mathcal R_n|/(-\Imn\omega_n)$ approaches a geometric quantity $\Gamma$, and that the overtone sum converges absolutely for
\begin{equation*}
 t-t'>\max(\Gamma_{\pr},\Gamma_{\ob}).
\end{equation*}
For each family of modes, the radial contribution to $\Gamma$ is the light-travel time from a bounce radius to the source radius plus the light-travel time from the same radius to the observer radius, along the special null rays identified in
Sec.~\ref{s:geometry} [Eqs.~\eqref{e:gpr} and~\eqref{e:gob}]. The bounce radii lie close to the black hole, at $r\simeq2.497M$ (oblate, $r_*\simeq0.204M$)
and $r\simeq2.507M$ (prolate, $r_*\simeq0.244M$) for $a/M=0.7$, and both reduce to the Schwarzschild value $r\simeq2.557M$ ($r_*=0$) as $a\to0$.
For prolate modes, which concentrate near the equatorial plane, 
the angular contribution subtracts $a(2-\sin\theta-\sin\theta')$ from the radial  part of $\Gamma_\pr$ when the source or observer lies off the equator; oblate modes have no such reduction. This angular term extends the prolate bounce-radius
picture to two angle-dependent bounce surfaces, the
zero-time sections of the null fronts. The oblate bounce
surface has a constant radius, while the associated spacetime
convergence boundary is timelike away from the rotation
axis at nonzero spin. In Schwarzschild, both constructions
recover the same bounce sphere. As a result, at $a/M=0.7$ the oblate modes set the convergence time whenever the source or observer lies beyond the prolate bounce, and the prolate modes can set it only when both points are close to the black hole and near the equator. The combined convergence threshold need not coincide
with the first causal arrival: at fixed angular indices,
the QNM sums can converge before any signal reaches
the observer.

We tested these predictions with amplitudes computed numerically, including the radial wave functions, the angular functions at both endpoints and the pole normalization. Their growth with damping reproduces $\Gamma_{\pr}$ and $\Gamma_{\ob}$ as functions of the observer position with maximum absolute percentage differences of $0.65\%$ and $0.91\%$, respectively, for the most damped pole pairs. Ratios of amplitudes of neighboring modes reproduce the predicted phases without fitting.

The condition above concerns the Green function, i.e., a single source point $(r',\theta')$ and a single emission time $t'$. For a physical source, such as a point-particle plunging into the BH, the field at the observer is a superposition of such impulses. Convolving the Green function with the source gives rise to dynamical QNM amplitudes~\cite{DeAmicis:2025xuh, DeAmicis:2026wqd}: our prescription for the bounce radius will allow us to generalize these computations to rotating BHs, extending the regime of validity of QNM models at earlier times closer to the peak of the waveform. It has recently been shown that the QNM start time, defined as the time at which a constant amplitude QNM model converges to the waveform of a plunging point-particle after summing over all overtones, occurs strictly after the bounce time for non-rotating BHs~\cite{Kuntz:2026crc}. It will be interesting to understand the interplay between the QNM start time and the bounce radius for rotating BHs too.  

\Needspace{8\baselineskip}
\begin{acknowledgments}
During the preparation of this manuscript, we became aware
of independent work by Arnaudo and Withers~\cite{awKerr2026},
who determine the Kerr bounce radii and convergence regions
of the radial part of the Green function for
the prolate and oblate families from complex-time singularities. Our growth rates coincide
with their radial ones at $\theta=\theta'=\pi/2$. 

E.B. and D.C. acknowledge funding from the European Union's Horizon ERC Synergy Grant ``Making Sense of the Unexpected in the Gravitational-Wave Sky'' (Grant No.~GWSky--101167314). A.K  thanks the Fundação para a Ciência e Tecnologia (FCT), Portugal, for the financial support to the Center for Astrophysics and Gravitation (CENTRA/IST/ULisboa) through grant No. UID/PRR/00099/2025 and grant No. UID/00099/2025, as well as to the FCT project ``Gravitational waves as a new probe of fundamental physics and astrophysics'' grant agreement 2023.07357.CEECIND/CP2830/CT0003. We used Claude (Anthropic; Claude Opus~5.5 and Claude Fable~5.1, via Claude Code) and Astra (OpenAI) to help check calculations, search and verify the literature, and edit the text. The authors reviewed and verified all derivations and results, and take full responsibility for the manuscript.
\end{acknowledgments}
\section*{Data availability}
The Mathematica and SymPy notebooks containing the analytic calculations, and the Python codes used for the numerical computations, will be made publicly available in a GitHub repository at \url{https://github.com/CiattagliaDavide/kerr-qnm-convergence} 
upon publication.

\appendix

\section{Angular matching and the projector norm}\label{a:angular}

Here we complete the polar and equatorial matching used
in Sec.~\ref{s:wkb}
and estimate the bilinear norm entering the angular projector.
We check whether normalization changes the exponential
rates in Eq.~\eqref{e:angularrates}.
For $s=-2$, $m=2$ and the branches in Eq.~\eqref{e:Abalances},
we normalize the prolate wave by
$S_n(\pi/2)=1$ and the oblate wave by $S_n(0)=1$.
In both cases the norm is bilinear,
$N_n=\int_0^\pi S_n(\theta)^2\sin\theta\,\dd\theta$.

\textit{Prolate.} Near $\theta=0$, on the scale
$\theta=O((-a\omega_{\rm I})^{-1})$, Eq.~\eqref{e:TeukS} becomes
\begin{equation*}
 \partial_\theta^2 S +\frac{1}{\theta} \partial_\theta S
 -(a\omega_{\rm I})^2S=0.
\end{equation*}
The equation above is the modified Bessel equation of order
zero. Its two independent solutions are
$I_0(-a\omega_{\rm I}\theta)$ and
$K_0(-a\omega_{\rm I}\theta)$, the modified Bessel functions
of the first and second kind. The function $I_0$ is regular with $I_0(0)=1$, whereas $K_0$ diverges
logarithmically there. Regularity at $\theta=0$ therefore
requires $S\propto I_0(-a\omega_{\rm I}\theta)$.

To connect this solution to WKB, we compare their exponential
dependence at small $\theta$ with
$(-a\omega_{\rm I})\theta\gg1$. These conditions are compatible
at large damping: the angle can still be small while the
Bessel argument is already large. In this limit~\cite{dlmfBessel},
\begin{equation*}
 I_0(-a\omega_{\rm I}\theta)
 \sim\frac{e^{-a\omega_{\rm I}\theta}}
 {\sqrt{2\pi(-a\omega_{\rm I})\theta}}.
\end{equation*}
Since $\sin\theta\simeq\theta$ near $\theta=0$, this selects
the WKB branch whose modulus grows as $\theta$ increases
towards $\pi/2$. 
Near $\theta=\pi$, regularity instead selects
$I_4[-a\omega_{\rm I}(\pi-\theta)]$, with the same growing
exponential towards the equator.
Relative to the equator, the WKB exponential selected by regularity at $\theta=0,\pi$ is therefore
$e^{-ia\omega(1-\sin\theta)}$, as in
Ref.~\cite{bertiAngular2006}.

We have selected the WKB solution on each side, but still
need to connect the two expressions through $\theta=\pi/2$,
where the leading potential vanishes and WKB fails because
of the $Q_\theta^{-1/4}$ prefactor
(cf. Eq.~\eqref{e:localradialwkb}).
Setting $y=\sqrt{-a\omega_{\rm I}}(\theta-\pi/2)$
and expanding Eq.~\eqref{e:angularrescaling} near the equator,
we find that the leading local equation and its matched
solution are~\cite{bertiAngular2006}
\begin{equation}
 \partial_y^2\widetilde S+(1-y^2)\widetilde S=0,
 \qquad \widetilde S\propto e^{-y^2/2}.
 \label{e:equatorialoscillator}
\end{equation}
This equation has two independent solutions, both regular at
$y=0$. We need the one that decays away from the equator to
match the WKB solution already selected. This is the Gaussian;
the other solution grows at large $|y|$. Since
$1-\sin\theta\simeq(\theta-\pi/2)^2/2$, the Gaussian matches
the selected exponential on both sides.

The prolate profile is concentrated near $\theta=\pi/2$.
Our normalization $S_n(\pi/2)=1$ fixes the Gaussian in
Eq.~\eqref{e:equatorialoscillator} to unit height.
Since $S_n^2\sin\theta=\widetilde S_n^2$, changing variables
to $y=\sqrt{-a\omega_{{\rm I},n}}(\theta-\pi/2)$ gives 
\begin{equation}
 N_n\sim\frac{1}{\sqrt{-a\omega_{{\rm I},n}}}
 \int_{-\infty}^{\infty}e^{-y^2}\,\dd y
 =\sqrt{\frac{\pi}{-a\omega_{{\rm I},n}}}.
 \label{e:weber}
\end{equation}
Dividing by this norm adds a factor proportional to
$\sqrt{-a\omega_{{\rm I},n}}$, with no exponential dependence.
At fixed angles away from $\theta=0,\pi$, we use
$S_n(\theta)=u(\theta)e^{-ia\omega_n(1-\sin\theta)}[1+o(1)]$
along our QNM sequence, with the same nonzero leading prefactor
$u(\theta)$ and $u(\pi/2)=1$.
This prefactor cancels in mode ratios at the same angles.

\textit{Oblate.} Equation~\eqref{e:obpolar} gives the regular
solution near $\theta=0$, following the expansion near that
pole in Ref.~\cite{casalsOttewill2005}. 
We already know the exponential factor. To estimate the norm,
we also need the factor multiplying it, so we write
\begin{equation*}
 S_n^{\mathrm{app}}(\theta)
 =f(\theta)e^{-a\omega_n(1-\cos\theta)}.
\end{equation*}
Substituting into Eq.~\eqref{e:TeukS}, the terms quadratic
in $\omega_n$ cancel. Setting the coefficient of the
linear terms to zero gives
\begin{equation*}
 \sin\theta\,f'+(1-\cos\theta)f=0,
 \qquad f(\theta)=C(1+\cos\theta).
\end{equation*}
The equation fixes the angular dependence but leaves
$C$ arbitrary. Our normalization $S_n(0)=1$ gives
$2C=1$, hence
\begin{equation}
 S_n^{\mathrm{app}}(\theta)
 =\frac{1+\cos\theta}{2}
 e^{-a\omega_n(1-\cos\theta)},
 \qquad S_n^{\mathrm{app}}(0)=1.
 \label{e:north}
\end{equation}
Near $\theta=0$, the prefactor tends to one, recovering
the normalized polar solution $e^{-v}$ of
Eq.~\eqref{e:obpolar}. 
Integrating its square by parts gives, for bounded
$\omega_{{\rm R},n}$,
\begin{equation}
 \int_0^\pi [S_n^{\mathrm{app}}(\theta)]^2
 \sin\theta\,\dd\theta
 =\frac{1}{2a\omega_n}
 +O(|a\omega_n|^{-2}).
 \label{e:obnorm}
\end{equation}
We must still estimate how correcting the profile near
$\theta=\pi$ changes the norm. We first determine the
width of this region, then bound the amplitude within it.

Write $\delta=\pi-\theta$. Using
$\sin\theta\simeq\delta$,
$\cos\theta\simeq-1+\delta^2/2$ and
$A_{\ob}=-(a\omega_n)^2-2a\omega_n+O(1)$,
Eq.~\eqref{e:TeukS} gives, at the relevant orders,
\begin{equation*}
 \partial_\delta^2 S+\frac{1}{\delta}\partial_\delta S
 -\left[(a\omega_n)^2\delta^2
 +6a\omega_n+\frac{16}{\delta^2}\right]S\simeq0.
\end{equation*}
The quadratic and singular terms become comparable when
\begin{equation*}
 |a\omega_n|^2\delta^2\sim\delta^{-2},
 \qquad
 \delta=O(|a\omega_n|^{-1/2}).
\end{equation*}
This sets the width of the region requiring a correction.
The term linear in $\omega_n$ also contributes on this scale.

To estimate the amplitude, set
$v=a\omega_n(1+\cos\theta)$.
The profile already found in Eq.~\eqref{e:north} gives
\begin{align*}
 S_n^{\mathrm{app}}
 &=\frac{e^{-2a\omega_n}}{2a\omega_n}\,ve^v,\\
 \partial_v S_n^{\mathrm{app}}
 &=\frac{e^{-2a\omega_n}}{2a\omega_n}(1+v)e^v.
\end{align*}
At a fixed large $|v|$ in the matching region, both
are $O(|a\omega_n|^{-1})$:
the factor $|e^{-2a\omega_n}|=e^{-2a\omega_{{\rm R},n}}$
remains bounded along our sequence.

In terms of $v$, the leading equation near $\theta=\pi$
becomes
\begin{equation*}
 v^2 \partial_v^2S+v \partial_v S-(v^2+3v+4)S=0.
\end{equation*}
This equation contains no frequency parameter.
Regularity fixes the shape of its solution up to an
overall factor. Matching its value and derivative to
the estimates above fixes the amplitude at
$O(|a\omega_n|^{-1})$, so the regular solution remains
of this order throughout a fixed range of $v$.

The integrated measure of this region is
$\int\sin\theta\,\dd\theta=O(\delta^2)
=O(|a\omega_n|^{-1})$.
Bounding the modulus of its contribution to the bilinear
norm by the maximum of $|S_n|^2$ times this measure gives
\begin{equation*}
 \underbrace{O(|a\omega_n|^{-2})}_{\text{squared amplitude}}
 \;\underbrace{O(|a\omega_n|^{-1})}_{\text{integrated measure}}
 =O(|a\omega_n|^{-3}).
\end{equation*}
The approximation in Eq.~\eqref{e:north} satisfies the
same bound there. Replacing it by the regular solution
therefore leaves the leading norm $1/(2a\omega_n)$
unchanged at this order.

The inverse norm consequently contributes only a power
of damping, leaving the exponential rates in
Eq.~\eqref{e:angularrates} unchanged.
Figure~\ref{f:angularprofiles} compares these rates with
the numerical projector, computed from the numerical
angular profiles and their bilinear norm.
\section{Schwarzschild limit}\label{a:schwarzschild} 
In this Appendix we take the zero-spin limit of the optical actions and growth rates introduced in Secs.~\ref{s:wkb} and~\ref{s:competition}. We recover the common Schwarzschild action, bounce radius and convergence condition.

As $a\to0$ at fixed $r, r'$, both radial slopes become
\begin{equation}
 D'_{\pr},D'_{\ob}\longrightarrow
 \frac{r}{r-2}=\frac{\dd r_*^{\rm Schw}}{\dd r}.
 \label{e:schwslopes}
\end{equation}
The two Kerr turning points used as origins of the actions,
$r_{\rm tp}$ and $ia$, both approach $r=0$, the Schwarzschild
turning point of the leading radial momentum.
We must also recover the integration constant: adding a
constant to an action leaves its radial slope unchanged
but can shift the bounce defined by its vanishing real part.
We therefore check that moving the Kerr integration origins
to zero leaves no finite contribution.

For prolate, we can write
\begin{equation*}
 z_{\pr}(r)
 =\int_0^r D'_{\pr}(u)\,\dd u
 -\int_0^{r_{\rm tp}}D'_{\pr}(u)\,\dd u.
\end{equation*}
The last integral is the correction due to the displaced
turning point, so we need it to vanish. 

The turning-point equation $r_{\rm tp}^3+a^2r_{\rm tp}+2a^2=0$
gives $|r_{\rm tp}|=O(a^{2/3})$.
Along the straight segment from zero to $r_{\rm tp}$,
$|\sqrt{q_0(u)}|=O(a\sqrt{|u|})$, while $|\Delta(u)|$
is bounded below by a constant times $|u|$.
Thus $|D'_{\pr}(u)|=O(a/\sqrt{|u|})$, and integrating
this bound gives
\begin{equation*}
 \left|\int_0^{r_{\rm tp}}D'_{\pr}(u)\,\dd u\right|
 =O\!\left(a\sqrt{|r_{\rm tp}|}\right)
 =O(a^{4/3})\longrightarrow0.
\end{equation*}
Thus moving the lower endpoint to zero leaves no finite
contribution. 
The integral of $D'_{\pr}-D'_{\ob}$ from zero to $r$ also
vanishes. Thus the prolate action approaches the
Schwarzschild tortoise coordinate measured from zero.
For oblate, the integration origin is explicit in
$z_{\ob}(r)=r_*(r)-r_*(ia)$, and
$r_*(ia)\to r_*^{\rm Schw}(0)=2\pi i$ since $\log(-1)=i\pi$
on this branch.
Subtracting this initial value yields
\begin{equation}
 z_{\pr},z_{\ob}\longrightarrow z_0(r)
 =r+2\log(r/2-1)-2\pi i.
 \label{e:schw}
\end{equation}
This is the Schwarzschild turning-point action used by
Oshita~\cite{oshita2026}. The bounce condition
$\Ren z_0(r_{\rm b}^{\rm Schw})=0$ gives
\begin{equation}
 r_{\rm b}^{\rm Schw}=2\big[1+W_0(e^{-1})\big]\simeq2.557,
 \label{e:schwbounce}
\end{equation}
where $W_0$ is the principal Lambert function.
Taking the real part, both signed optical distances therefore become
\begin{equation}
 \begin{aligned}
 D_{\pr}(r),D_{\ob}(r)&\longrightarrow D_0(r),\\
 D_0(r)&=r_*^{\rm Schw}(r)-r_*^{\rm Schw}(r_{\rm b}^{\rm Schw}).
 \end{aligned}
 \label{e:schwaction}
\end{equation}
The angular term vanishes, and the two rates become
\begin{equation}
 \Gamma_{\pr},\Gamma_{\ob}\longrightarrow
 \Gamma_0=|D_0(r)|+|D_0(r')|.
 \label{e:schwrate}
\end{equation}
Equivalently,
\begin{equation*}
 \begin{aligned}
 \Gamma_0=\max\big\{&|D_0(r)-D_0(r')|,\\
                   &|D_0(r)+D_0(r')|\big\}.
 \end{aligned}
\end{equation*}
Thus $t-t'>\Gamma_0$ is the Schwarzschild condition of
Refs.~\cite{Arnaudo:2025uos,Arnaudo:2026tcy}, with their tortoise
constant $C=r_*^{\rm Schw}(r_{\rm b}^{\rm Schw})$.
For endpoints at the same angular position, this equals the direct
radial travel time on opposite sides of the bounce. On the same side,
it includes an extra round trip to the bounce from the nearer endpoint.
\section{Optical geometry}\label{a:optical_geometry}  We derive the causal bounds used in Sec.~\ref{s:competition} and distinguish the motion of a front at fixed polar angle from the motion of its null generators.

To obtain causal bounds, define $\mathcal A=(r^2+a^2)^2-a^2\Delta\sin^2\theta$. Completing the square in the azimuthal part of the metric gives
\begin{equation}
 \begin{aligned}
 \dd s^2={}&-\frac{\Sigma\Delta}{\mathcal A}\dd t^2
 +\frac{\Sigma}{\Delta}\dd r^2+\Sigma\dd\theta^2\\
 &+\frac{\mathcal A\sin^2\theta}{\Sigma}
 \left(\dd\phi-	\frac{2ar}{\mathcal A}\dd t\right)^2.
 \end{aligned}
 \label{e:app_squaremetric}
\end{equation}
For $r>r_+$, the spatial terms are nonnegative. For a future-directed causal curve, retaining the radial term alone bounds the radial coordinate speed, $|\dd r/\dd t|\le\Delta/\sqrt{\mathcal A}$. At a fixed radius this upper bound is largest on the equator and smallest on the axis; the two rays in Eq.~\eqref{e:rays} attain those respective bounds.

Retaining the polar term instead gives
\begin{equation}
 \dd t\ge\sqrt{\mathcal A/\Delta}\,|\dd\theta|
       \ge r_+|\dd\theta|.
 \label{e:app_frontcausal}
\end{equation}
The second inequality follows from
\begin{equation*}
 \mathcal A-r^2\Delta
 =2r(r^2+a^2)+a^2\Delta\cos^2\theta>0
\end{equation*}
and $r>r_+$. Integrating gives $t-t'\ge r_+|\theta-\theta'|$, even when the curve changes polar direction. For the example in Sec.~\ref{s:competition}, $a/M=0.7$ and $|\theta-\theta'|=\pi/3$, this lower bound is $1.795M$.

Finally, the motion of a front at fixed $\theta$ differs from the motion of its null generators. For $t=T_\pm$, the fixed-angle intersection obeys $\dd r/\dd t=\pm\Delta/\sqrt{q_0}$. The future-directed generators are proportional to $-g^{\mu\nu}\partial_\nu(t-T_\pm)$ and instead obey
\begin{equation}
 \left.\frac{\dd r}{\dd t}\right|_{\rm ray}
 =\pm\frac{\Delta\sqrt{q_0}}{\mathcal A},
 \qquad
 \left.\frac{\dd\theta}{\dd t}\right|_{\rm ray}
 =\frac{a\Delta\cos\theta}{\mathcal A}.
 \label{e:app_generators}
\end{equation}
The two radial velocities agree on the equator. Away from it, the generators also move in polar angle. The horizontal arrows in Fig.~\ref{f:vi_snap} track front intersections, not these ray trajectories.

\section{Numerical calculation and tests of the amplitudes}
\label{a:numerical}\label{s:num}

This appendix contains details on our numerical results.
We first describe how we compute the complete amplitudes in
Eq.~\eqref{e:mainres}. We then compare their growth rates with
WKB, estimate the subleading powers of damping and test the
complex amplitude ratios. The growth-rate tests include the pole
normalization and test whether the reference amplitude in
Eq.~\eqref{e:Rb} adds any further exponential dependence on damping.

\subsection{Computing the modes and amplitudes}\label{s:numericalmethod}

We compute the prolate and oblate sequences for $s=-2$,
$m=2$ and $a/M=0.7$ in the following steps.

\medskip
\noindent\textit{1. Trial frequency and angular solution.}
We start from estimates guided by the known large-damping behavior
of the frequencies~\cite{keshet2008,keshetBenMeir2014} and angular
eigenvalues in Eq.~\eqref{e:Abalances}. As a separate check, we used a Python solver to follow
selected modes from Schwarzschild to $a/M=0.7$, tracking
their frequencies and angular eigenpairs as the spin increased.
We reached at least $n=500$ for $(\ell,m)=(2,2)$ and
$n=350$ for $(3,2)$. At high damping, their angular
eigenvalues agree with the prolate and oblate asymptotic
expansions, respectively. The code, data and comparisons
will be made available in the accompanying repository.
For each trial frequency, we compute $A(\omega)$ by solving the
angular equation in a truncated basis of spin-weighted spherical
harmonics. This basis already enforces regularity at $\theta=0,\pi$.
The eigenvector gives the angular profile and its bilinear norm,
which enter the projector in Eq.~\eqref{eq:angularProj}.

\medskip
\noindent\textit{2. Radial solution.}
For each $(\omega,A)$, we evaluate the horizon-ingoing solution
through its confluent-Heun representation~\cite{Fiziev2010}:
\begin{equation}
 \begin{aligned}
 R_{\IN}(r)\propto{}&e^{-i\omega r}
 (r-r_+)^{-s-i\eta_+}(r-r_-)^{-s+i\eta_-}H(x),\\
 x={}&\frac{r_+-r}{r_+-r_-},\qquad
 \eta_\pm=\frac{2r_\pm\omega-am}{r_+-r_-}.
 \end{aligned}
 \label{e:numheun}
\end{equation}
The overall factor is fixed by the unit horizon amplitude in
Eq.~\eqref{eq:RH_asymp}. For the numerical calculation, we set this
factor to one and use the same normalization for the radial
amplitudes and Wronskian. This rescaling cancels from the complete
amplitude at a QNM pole. 
Substitution into Eq.~\eqref{e:TeukR} gives
\begin{equation}
 H''+\left(\frac{\gamma}{x}+\frac{\delta}{x-1}
 +\epsilon\right)H'
 +\frac{\alpha x-q_H}{x(x-1)}H=0,
 \label{e:numheunode}
\end{equation}
where derivatives are with respect to $x$, and
\begin{equation}
 \begin{aligned}
 \epsilon&=2i\omega(r_+-r_-),&
 \alpha&=(1-2s)\epsilon,\\
 \gamma&=1-s-2i\eta_+,&
 \delta&=1-s+2i\eta_-,\\
 q_H&=\lambda+2s+2i(1-2s)r_+\omega.
 \end{aligned}
 \label{e:numheunparameters}
\end{equation}
We use the solution regular at $x=0$, normalized by $H(0)=1$.
This determines the radial profile and its derivative.

\medskip
\noindent\textit{3. Amplitudes at infinity and QNM frequencies.}
We construct the asymptotic series
\begin{equation}
 \begin{aligned}
 R_{\infty}^{\rm in}&\sim r^{-1}e^{-i\omega r_*}
 [1+O(r^{-1})],\\
 R_{\infty}^{\rm out}&\sim r^3e^{i\omega r_*}
 [1+O(r^{-1})],
 \end{aligned}
 \label{e:nummatching}
\end{equation}
with coefficients obtained from the radial equation.
Writing
\begin{equation}
 R_{\IN}=B_{\rm in}R_{\infty}^{\rm in}
          +B_{\rm out}R_{\infty}^{\rm out},
 \label{e:numamplitudes}
\end{equation}
and differentiating with respect to $r$ gives a $2\times2$ linear
system, which we solve for $B_{\rm in}$ and $B_{\rm out}$.
We repeat steps 1--3, refining the trial frequency to impose
$B_{\rm in}=0$ numerically. This determines the QNM frequency
and the corresponding angular eigenvalue and profile.

\medskip
\noindent\textit{4. Complete amplitudes.}
With $R_{\UP}$ normalized to unit outgoing amplitude, the
Wronskian and its derivative at a simple QNM pole satisfy
\begin{equation}
 \widehat W=2i\omega B_{\rm in},\qquad
 \widehat W'(\omega_n)=2i\omega_n B_{\rm in}'(\omega_n).
 \label{e:numwronskian}
\end{equation}
The derivatives are total derivatives with respect to $\omega$,
taking into account the $\omega$-dependence of the separation
constant $A(\omega)$. They are evaluated numerically.
We then assemble Eq.~\eqref{e:mainres}:
\begin{equation}
 \mathcal R_n=
 \frac{\Pi_n(\theta,\theta')\,
 R_{\IN,n}(r_<)R_{\UP,n}(r_>)}
 {\widehat W'(\omega_n)}.
 \label{e:numamplitude}
\end{equation}
At the pole, $R_{\UP,n}=R_{\IN,n}/B_{{\rm out},n}$, so both radial
factors are obtained from the same Heun solution.
The Wronskian derivative is computed once per pole and reused
for all source and observer positions.

\medskip
The precision and angular basis size increase with damping.
The radial truncation order is chosen from the expected
large-order behavior of the series.
Implementation details will accompany the public code described
under Data availability.

\begin{figure*}[!tbp]
\centering
\includegraphics[width=.98\textwidth]{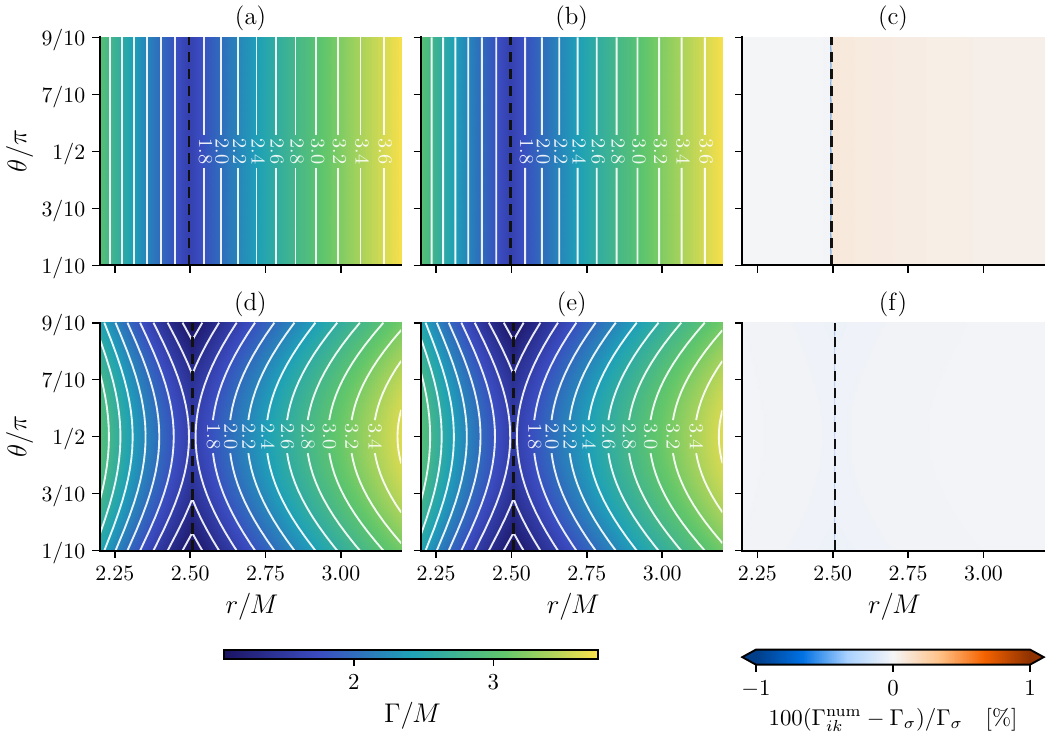}
\caption{\label{f:maps}WKB rates $\Gamma_\sigma/M$ (left),
numerical rates $\Gamma^{\rm num}_{ik}/M$ [Eq.~\eqref{e:secant}]
(middle), and percentage differences
$100(\Gamma^{\rm num}_{ik}-\Gamma_\sigma)/\Gamma_\sigma$ (right).
Here $a/M=0.7$ and the source is at $(3M,\pi/2)$.
Top: oblate, $(i,k)=(8000,16000)$.
Bottom: prolate, $(i,k)=(4000,6000)$.
The first two columns share a color scale.
White contours mark constant rates; dashed lines mark the bounces.
The percentage color scale runs from $-1\%$ to $1\%$.}
\end{figure*}

\subsection{Growth rates}
At finite damping, the magnitude of one amplitude contains both
exponential growth and prefactors. At fixed source and observer
positions, the expected form is
\begin{equation}
 \log|\mathcal R_{\sigma,n}|\simeq
 -\Gamma_\sigma\omega_{{\rm I},n}
 +p_\sigma\log(-M\omega_{{\rm I},n})+c+\cdots.
 \label{e:finitegrowth}
\end{equation}
Here $p_\sigma$ is the power of damping in the complete amplitude
and $c$ is independent of overtone number.
To isolate the exponential rate, we measure the slope between
two modes of the same sector,
\begin{equation}
 \Gamma^{\rm num}_{ik}=
 \frac{\log|\mathcal R_k|-\log|\mathcal R_i|}
 {\omega_{{\rm I},i}-\omega_{{\rm I},k}},
 \qquad \omega_{{\rm I},k}<\omega_{{\rm I},i}.
 \label{e:secant}
\end{equation}
The difference removes $c$. The power contributes
$p_\sigma\log[(-\omega_{{\rm I},k})/(-\omega_{{\rm I},i})]$
divided by the damping separation. This correction tends to zero
as both dampings increase. Thus we can compare this slope directly
with $\Gamma_\sigma$ without fitting an amplitude or power.

Figure~\ref{f:maps} compares the predicted and measured rates
for different observer positions, with the source fixed.
Since the numerical amplitudes include the Wronskian derivative,
agreement tests the complete combination of radial profiles,
angular projector and pole normalization.
On the displayed grid, the largest absolute percentage differences
occur near the bounces: $0.91\%$ for oblate and $0.65\%$ for prolate.
The change of dominant radial contribution there makes
finite-damping corrections particularly relevant.

\subsection{Subleading prefactors}\label{a:model}
To determine $p_\sigma$, we remove the predicted exponential growth
from the numerical amplitudes and fit the remaining dependence on
damping. At fixed source and observer positions above both bounce
radii, we use
\begin{equation}
 \begin{aligned}
 \log|\mathcal R_{\sigma,n}|+\Gamma_\sigma\omega_{{\rm I},n}
 \simeq{}&c+p_\sigma\log(-M\omega_{{\rm I},n})\\
 &+\frac{b}{-M\omega_{{\rm I},n}}.
 \end{aligned}
 \label{e:powerfit}
\end{equation}
Here $c$ sets the overall magnitude and $b$ gives the leading
inverse-damping correction within this fit model.
We keep $\Gamma_\sigma$ at its WKB value and determine $c,p_\sigma,b$
by linear least squares, with equal weight for each mode in the fit.

\begin{figure}[!htbp]
\centering
\includegraphics[width=.98\columnwidth]{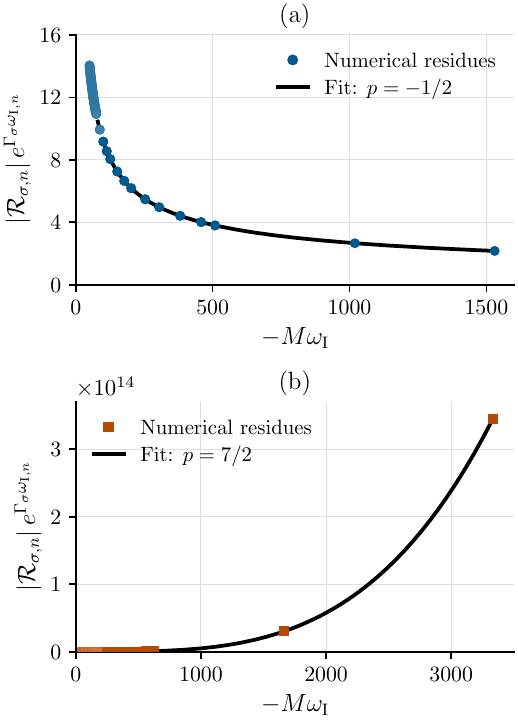}
\caption{\label{f:powers}\label{f:obpower}Amplitude magnitudes after
removing the WKB exponential, at $a/M=0.7$.
Points: numerical data. Curves: fits of Eq.~\eqref{e:powerfit},
dashed outside the fitted intervals.
The source is at $(2.6M,\pi/2)$; the observer is at
$(2.8M,\pi/2)$ for prolate (a) and $(3.2M,\pi/2)$ for oblate (b).}
\end{figure}

Figure~\ref{f:powers} shows the numerical magnitudes after removing
the WKB exponential, together with the exponentiated fits of
Eq.~\eqref{e:powerfit}. Both panels use linear axes and show the
same quantity. The fits give $p_{\pr}=-0.5006$ over
$100\le-M\omega_{\rm I}\lesssim1530$ and $p_{\ob}=3.4999$ over
$240\le-M\omega_{\rm I}\lesssim3333$. They use 24 prolate modes
and 20 oblate modes, respectively. These finite-damping results
support the simple values $-1/2$ and $7/2$ used below.

\subsection{Complex amplitude ratios}\label{s:complex}
The growth-rate test uses amplitude magnitudes. We now compare complex
ratios, which also test the phase predicted by WKB. For these tests,
both radii lie above the sector's bounce, where the plus exponential
gives the leading radial contribution (Sec.~\ref{s:wkb}).

We retain the full action $z_\sigma=D_\sigma+i\beta_\sigma$.
In $e^{i\omega z_\sigma}=e^{i\omega D_\sigma}e^{-\omega\beta_\sigma}$,
the second factor has bounded modulus for bounded $\omega_{\rm R}$,
but its phase changes with damping. Combining the radial actions with
the angular profiles gives the optical factor $e^{i\omega t_\sigma}$,
where
\begin{align}
 t_{\pr}&=z_{\pr}(r)+z_{\pr}(r')
             -a(2-\sin\theta-\sin\theta'),\label{e:opticalpr}\\
 t_{\ob}&=z_{\ob}(r)+z_{\ob}(r')
             +ia(2-\cos\theta-\cos\theta').\label{e:opticalob}
\end{align}
At these radii, $\Ren t_\sigma=\Gamma_\sigma$.
Since the phase of $e^{i\omega t_\sigma}$ is
$\Ren(\omega t_\sigma)$, the optical prediction for two modes is
\begin{equation}
 \arg\frac{\mathcal R_k}{\mathcal R_i}
 \simeq\Ren\!\left[(\omega_k-\omega_i)t_\sigma\right]
 \pmod{2\pi}.
 \label{e:opticalphase}
\end{equation}
Common leading angular factors cancel at fixed angles. Any remaining
mode-dependent phase in the normalization contributes to the
discrepancy.

\begin{figure}[!tbp]
\centering
\includegraphics[width=.48\textwidth]{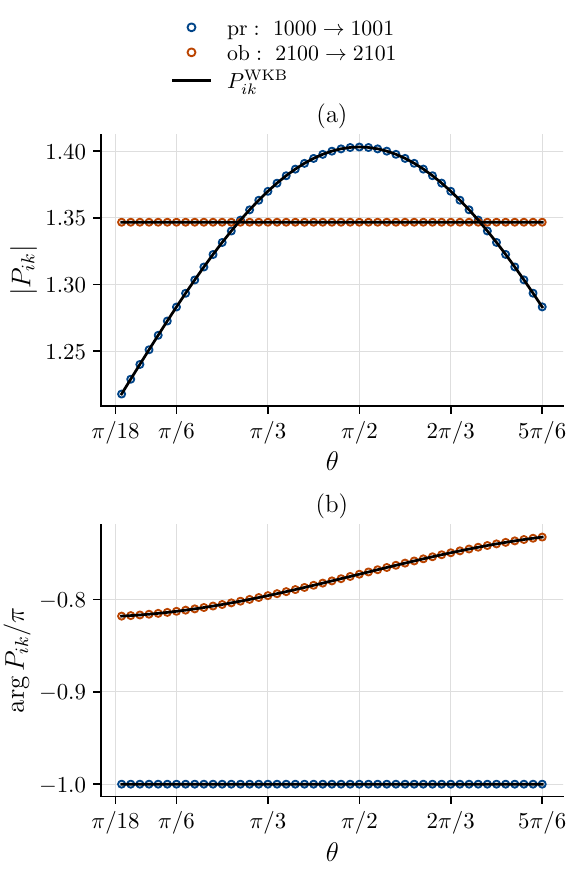}
\caption{\label{f:complex}Complete amplitude ratios at $a/M=0.7$:
magnitude (a) and phase in units of $\pi$ (b). The source is at
$(2.6M,\pi/2)$; the observer angle varies at $r=2.8M$.
Blue points: prolate $1000\to1001$. Orange points: oblate
$2100\to2101$. Points are sampled at $\theta=j\pi/60$,
$j=4,\ldots,50$. Curves show Eq.~\eqref{e:complexratio}, with
$p_{\pr}=-1/2$ and $p_{\ob}=7/2$ suggested by the separate power
fits. No amplitude or phase offset is fitted to these ratios.}

\end{figure}

To compare the modulus as well, we include a real power $p_\sigma$
of damping. We distinguish the ratio of numerical amplitudes from its
WKB prediction:
\begin{equation}
 \begin{aligned}
 P^{\rm num}_{ik}&=\frac{\mathcal R_k}{\mathcal R_i},\\
 P^{\rm WKB}_{ik}&=
 \left(\frac{-\omega_{{\rm I},k}}{-\omega_{{\rm I},i}}\right)^{p_\sigma}
 e^{i(\omega_k-\omega_i)t_\sigma}.
 \end{aligned}
 \label{e:complexratio}
\end{equation}
We use $p_{\pr}=-1/2$ and $p_{\ob}=7/2$, suggested by the fits
above. The power contributes a positive real factor, so it affects
only the modulus; the phase comparison is independent of it.
Figure~\ref{f:complex} compares $P^{\rm num}_{ik}$ and
$P^{\rm WKB}_{ik}$ as functions of the observer angle.
No amplitude or phase offset is fitted to these curves.

\subsection{An idealized pole sum}

At fixed source and observer positions, denote by $C_\sigma$
the complex prefactor remaining after extracting the power
of damping and the optical factor $e^{i\omega_n t_\sigma}$
from the amplitude. As a mathematical aside, we consider an
idealized model in which this prefactor is independent of
overtone number. Including the time dependence gives
\begin{equation}
 \begin{aligned}
 e^{-i\omega_n(t-t')}\mathcal R_{\sigma,n}
 \simeq{}&C_\sigma(-M\omega_{{\rm I},n})^{p_\sigma}\\
 & e^{-i\omega_n[(t-t')-t_\sigma]}.
 \end{aligned}
\end{equation}
Using the asymptotic spacing
$\omega_n\simeq\omega_*+in\Delta\omega_{\rm I}$,
$\Delta\omega_{\rm I}<0$~\cite{keshet2008,keshetBenMeir2014},
the power of damping becomes proportional to $n^{p_\sigma}$ at
leading order, and the exponential becomes a fixed factor raised
to $n$. The remaining factors are independent of $n$ and come
outside the sum. We can therefore use the defining series~\cite{dlmfPoly}
\begin{equation}
 \Li_s(z)=\sum_{n=1}^{\infty}\frac{z^n}{n^s},\qquad |z|<1,
 \label{e:polylogdefinition}
\end{equation}
with $s=-p_\sigma$, obtaining
\begin{equation}
 \begin{aligned}
 &\sum_{n=1}^{\infty}n^{p_\sigma}
 \left\{e^{\Delta\omega_{\rm I}[(t-t')-t_\sigma]}\right\}^{n}\\
 &\qquad=\Li_{-p_\sigma}\!\left(e^{\Delta\omega_{\rm I}[(t-t')-t_\sigma]}\right).
 \end{aligned}
 \label{e:mainli}
\end{equation}
The argument has modulus below one when
$t-t'>\Ren t_\sigma=\Gamma_\sigma$.
Starting the idealized sum at a higher overtone subtracts finitely
many terms. The proposed powers give $\Li_{1/2}$ for prolate and
$\Li_{-7/2}$ for oblate.

\clearpage

\bibliographystyle{apsrev4-2}
\bibliography{bibliografia}
\end{document}